\documentclass[12pt]{article}   
\usepackage{amssymb,undertilde} 

\usepackage{amsmath,amscd} 

\begin{document}


\title{The Nontriviality of Trivial General Covariance: How  Electrons Restrict `Time' Coordinates,  Spinors (Almost) Fit into Tensor Calculus, and $\frac{7}{16}$ of a Tetrad Is Surplus Structure\footnote{Forthcoming in \emph{Studies in History and Philosophy of Modern Physics} }} 
\author{J. Brian Pitts \\ Department of Philosophy, Department of Physics, and \\ Reilly Center for Science, Technology and Values \\University of Notre Dame\\ Notre Dame, Indiana 46556 USA }

\maketitle 
\begin{abstract}

It is a commonplace in the philosophy of physics that any local physical theory can be represented using arbitrary coordinates, simply by using tensor calculus.  On the other hand, the physics literature often claims  that spinors \emph{as such} cannot be represented in coordinates in a curved space-time. These commonplaces are inconsistent.  What  general covariance means for theories with fermions, such as electrons, is thus unclear. In fact both commonplaces are wrong. Though it is not widely known, Ogievetsky and Polubarinov constructed spinors in coordinates in 1965, enhancing the unity of physics and helping to spawn particle physicists' concept of nonlinear group representations. Roughly and locally, these spinors resemble the orthonormal basis or ``tetrad'' formalism in the symmetric gauge, but they are conceptually self-sufficient and more economical.  The typical tetrad formalism is de-Ockhamized, with six extra field components and  six compensating gauge symmetries to cancel them out. The Ogievetsky-Polubarinov formalism, by contrast, is (nearly) Ockhamized, with most of the fluff removed.  As developed nonperturbatively by Bilyalov, it admits any coordinates at a point, but ``time'' must be listed first.  Here ``time'' is defined in terms of an eigenvalue problem involving the metric components and the matrix $diag(-1,1,1,1)$, the product of which must have no negative eigenvalues in order to yield a real symmetric square root that is a function of the metric.  Thus  even  formal general covariance requires reconsideration; the atlas of admissible coordinate charts should be sensitive to  the types and \emph{values} of the fields involved.

Apart from coordinate order and the usual spinorial two-valuedness, (densitized) Ogievetsky-Polubarinov spinors form, with the (conformal part of the) metric, a nonlinear geometric object, for which  important results on Lie and covariant differentiation are recalled. 
Such  spinors avoid a spurious absolute object in the Anderson-Friedman analysis of substantive general covariance.  They also permit the gauge-invariant localization of the infinite-component gravitational energy in General Relativity.  Density-weighted spinors exploit the conformal invariance of the massless Dirac equation  to show that the volume element is absent. Thus instead of an arbitrary nonsingular matrix with 16 components, 6 of which are gauged away by a new local $O(1,3)$ gauge group and one of which is irrelevant due to conformal covariance, one can, and presumably should, use density-weighted Ogievetsky-Polubarinov spinors coupled to the 9-component symmetric square root of the part of the metric that fixes null cones. Thus $\frac{7}{16}$ of the orthonormal basis is eliminated as surplus structure.  Greater unity between spinors (related to fermions, with half-integral spin) and tensors and the like (related to bosons, with integral spin) is achieved, such as regarding conservation laws.

Regarding the conventionality of simultaneity, an unusually wide range of $\epsilon$ values is admissible, but some extreme values are inadmissible.  Standard simultaneity uniquely makes the spinor transformation law linear and independent of the metric, because transformations among the standard Cartesian coordinate systems fall within the conformal group, for which the spinor transformation law is linear.   The surprising mildness of the restrictions on coordinate order as applied to the Schwarzschild solution is exhibited.

\end{abstract}  %



\section{Introduction} 

\subsection{A No-Go  Theorem from \emph{c.}  1930 Circumvented in the 1960s---But Who Noticed?}
Around the 1930s there arose in the  mathematical physics community a  no-go theorem pertaining to quantum mechanics, to the effect that there cannot be a treatment of certain distinctive of features of quantum mechanics in ways familiar from classical physics, not even classical relativistic field theory.  In the mid-1960s, with some clues in the early 1950s, the irrelevance of the theorem was shown by explicit construction of a formalism that did basically what the no-go theorem allegedly had shown to be impossible; some of the theorem's  technical assumptions were seen to be much less evident than they had seemed.  And yet great masses of literature went on with business as usual for decades, as though the progress at mid-century never happened. I do not have in mind to discuss the names of  Von Neumann \cite{VonNeumann} as proving a no-hidden-variables theorem around 1930,  Bohm \cite{Bohm} as showing a way around it in the early 1950s, and  Bell as providing systematic treatment of  the issue in the mid-1960s \cite{Bell}, as the reader might have assumed.  I have in mind another story involving  theorem involving 
Weyl \cite{WeylGravitationElectron,WeylElektronGravitation,WeylRice},  Fock \cite{FockDirac2}, Infeld and van der Waerden \cite{InfeldvanderWaerden}, and Cartan \cite{CartanSpinor} in the period 1929-37, with Weyl and Cartan stating a no-go theorem with special clarity and force;  Bryce  and C\'{e}cile DeWitt as providing clues regarding another way in the early 1950s \cite{DeWittSpinor}; 
 and substantial clarification and resolution in the mid-1960s by V. I. Ogievetsky\footnote{This closing vowels of this name are transliterated in different ways.  I use a ``y'' for convenience when no single specific paper is in view, while retaining the spelling actually used by each citation in the bibliography.} and I. V. Polubarinov \cite{OPspinorReprint,OP,OPspinorConf,BorisovOgievetskii,BilyalovConservation,Schucker,BilyalovSpinors,Iochum}. Like Bell's,  this clarification in the 1960s was not, from a sociological standpoint, by any means an immediate success.  Unlike the hidden variable issue, the question of the possibility of spinors as such in coordinates is \emph{still} shrouded in confusion and neglect.  Resolving that problem motivates this paper.

The no-go theorem in question is to the effect that \emph{there are no finite-component spinor fields with linear spinorial transformation laws under arbitrary coordinate transformations} 
 \cite{WeylGravitationElectron,WeylRice,CartanSpinor,GelfandLorentzSpinor}; it is also tacitly assumed that the spinor transformation law ought not to involve any other fields.  
This no-go theorem, with the tacit qualification made explicit, evidently is true.\footnote{Perhaps a few other premises too obvious to mention, such as that the spinor not vanish everywhere, are also involved. I aim here only for a physical, not mathematical, level of rigor. One theme of this work is the sometimes illusory character of the perfection suggested by precise modern mathematics. One needs to keep checking the intended domain of phenomena described to ascertain the appropriateness of the axioms.  } 
That the theorem does not have the significance that it seems to have will become evident after dropping the tacit exclusion of other fields, hedging on linearity (retaining linear dependence on the spinor but admitting nonlinear dependence on the other field, namely, the metric), and hedging a little on the arbitrariness of the coordinate transformations.
  At  times imprecision and failure to be aware of Ogievetsky-Polubarinov spinors has lead to strictly false assertions.  More formal mathematical works are more likely to start with truth and end with false glosses, rather than starting with falsehood.  Weyl's statements below mix truth and falsehood. It would be tedious to intervene sentence by sentence to separate the true from the false on every occasion, but below I will do that with the 1958 work of Gel'fand, Minlos and Shapiro \cite{GelfandLorentzSpinor} for illustration.

The no-go  theorem in question is \emph{usually taken to show that one cannot have spinors in a curved space-time  with spinorial behavior under coordinate transformations}.  Hence one defaults (typically) to the orthonormal basis formalism, in which spinors are coordinate \emph{scalars} and behave spinorially with respect to a new local $O(1,3)$ group of rotations and boosts of the orthonormal basis.\footnote{Sometimes one sees formalisms, whether old or new, that profess to do without a tetrad by adding even more surplus structure and an even larger gauge group to cancel it out. Belinfante's critique of such works, including Schr\"{o}dinger's, seems apt \cite{BelinfanteTensorUndor}.  It isn't very surprising, or very interesting, that one can devise a formalism with additional surplus structure and hide the tetrad in it.  For my purposes one succeeds in doing without a tetrad only if there is nothing equivalent to a tetrad in the formalism; certainly there should not be additional junk.}  
 This new gauge group reflects by how much the content of an orthonormal basis exceeds that of the metric tensor to which it corresponds. (The orthonormal basis is often called a vierbein or tetrad in four dimensions.)  
  Fock points out that an orthonormal basis is necessary for his ``half-vector'' formalism \cite{FockDirac2}.  Weyl is very perspicuous, not limiting his claim to those formalisms presently known while admitting room for future innovation.  All possible formalisms for spinors do not fit within tensor calculus, or (presumably) any sensible generalization thereof, he evidently holds:
\begin{quote}The components of [the tetrad] ${\bf e}(\alpha)$ in this co\"{o}rdinate system are designated by $e^p(\alpha).$  We need such local cartesian [sic] axes ${\bf e}(\alpha)$ in each point $P$ in order to be able to describe the quantity $\psi$ by means of its components $\psi^{+}_1$, $\psi^{+}_2$; $\psi^-_1,$ $\psi^-_2,$ for the law of transformation of the components $\psi$ can only be given for orthogonal transformations as it corresponds to a representation of the orthogonal group which cannot be extended to the group of all linear transformations.  The tensor calculus is consequently an unusable instrument for considerations involving the $\psi.$ \cite{WeylGravitationElectron} \end{quote}
In that same year Weyl explained that
\begin{quote}
The tensor calculus is not the proper mathematical instrument to use in translating the quantum-theoretic equations of the electron over into the \emph{general theory of relativity}.  Vectors and terms [tensors?] are so constituted that the law which defines the transformation of their components from one Cartesian set of axes to another can be extended to the most general linear transformation, to an affine set of axes.  That is not the case for quantity $\psi$, however; this kind of quantity belongs to a representation of the rotation group which cannot be extended to the affine group.  Consequently we cannot introduce components of $\psi$ relative to an arbitrary coordinate system in general relativity as we can for the electromagnetic potential and field strengths.  We must rather describe the metric at a point $P$ by local Cartesian axes $e(\alpha)$ instead of by the $g_{pq}.$ \cite{WeylRice} \end{quote}  
He has similar remarks in the second edition of the book on group theory and quantum mechanics \cite[p. 219]{WeylGroups}.

Cartan dramatically ends his book with the following 
\begin{quote}
THEOREM.  With the geometric sense we have given to the word ``spinor'' it is impossible to introduce fields of spinors into the classical Riemannian technique; that is, having chosen an arbitrary system of co-ordinates $x^i$ for the space, it is impossible to represent a spinor by any finite number $N$ whatsoever, of components $u_{\alpha}$ such that the $u_{\alpha}$ have covariant derivatives of the form 
$$ u_{\alpha.i} = \frac{ \partial u_{\alpha} }{\partial x^i } + \Lambda^{\beta}_{\alpha i}u_{\beta},$$
where the $\Lambda^{\beta}_{\alpha i}$ are determinate functions of $x^h$ [Footnote referring to \cite{InfeldvanderWaerden} and italics for most of the theorem have been suppressed.]. \cite[p. 151]{CartanSpinor} 
\end{quote} 
Such a result seems very decisive.  Yet it is ultimately misleading, because key premises that do the work of excluding Ogievetsky-Polubarinov spinors (if one may speak proleptically) are not even stated in the theorem.


As Scholz notes, the impossibility of having spinors in coordinates and consequent necessity of the use of an orthonormal basis was  recognized by its inventors as a \emph{conceptual innovation} \cite{ScholzWeylFockSpinor}.  Clearly they did not hedge their bets by presenting their results modestly as technical results subject to interpretive qualification by possible future technical developments. Had they done so, they would be blameless despite subsequent developments.  While no one should be faulted (in non-Whiggish history) for failing to prophesy the future of mathematical physics, more modest interpretations of results do have the virtue of leaving one less exposed to future refutation from unimagined quarters.  While the theorem that there is no \emph{linear} spinor transformation involving arbitrary coordinates is quite correct, its significance is less than  Weyl, Cartan and their many followers have thought.  On the contrary, the ambition of authors such as Schr\"{o}dinger \cite{SchrodingerDiracSpinor} to avoid a tetrad in coupling spinors to gravity were belatedly vindicated in an adequate formalism without some other surplus structure, as Ogievetsky and Polubarinov point out.

In fact Weyl's and Cartan's  conceptual innovation of the inadequacy of tensor calculus, founded as it was on incomplete technical results involving  an arbitrary tacit premise that would be exposed by OP in 1965,  \emph{was wrong}.  It is an artifact of the state of mathematical physics in the 1930s, as opposed to  developments in the 1950s and 1960s involving George Mackey's induced group representations  (though connection between the literature involving Ogievetsky-Polubarinov spinors and such pure mathematics was slow in coming), 
 nonlinear geometric objects in differential geometry, nonlinear group representations in particle physics, and the like---various more or less independent parallel lines of development.  Yet somehow  the main body of differential geometry has persevered with the 1930s tetrad technology  nonetheless, expressing  it in more modern-\emph{looking} form (\emph{e.g.}, \cite{SpinGeometry}), in terms of bundles, various words ending in ``morphism,'' and the like.  It therefore now \emph{looks} more advanced than most presentations of the OP formalism.  (Indeed it is more advanced, as far as extant treatment of global issues is concerned. It is noteworthy that the OP symmetric square root of the metric exists wherever the metric does; it suffers no topological obstruction.)  But that is largely  an illusion created by comparing apples and oranges, which would quickly disappear in comparing Ogievetsky and Polubarinov to Weyl and Cartan. The fact remains that the orthonormal basis formalism that forms the backbone of much modern global work on spinors in differential geometry is built on what was known and not known in the 1930s, some of which became dated in the 1950s and 1960s.   The sociological fact of the perseverance of obsolete results due to historical accidents is, clearly, not in itself a good reason for retaining those results. Of course there could be excellent reasons for working the OP formalism using the modern style.  This paper is premised partly on the modest idea that one should be able to walk before trying to run.  
Whereas tensor calculus was used and diffused for decades in component form in many books, the rather more demanding formalism of nonlinear geometric objects and the like, including OP spinors, has never had a pedagogical presentation.  And modern treatments are not especially easy to produce or to relate to classical ones \cite[p. v]{Spivak1} \cite[p. 159]{EarmanWorld}. This component treatment should facilitate a modern treatment by making OP spinors known and intelligible.  Some reasons why a `coordinate-free' treatment of nonlinear geometric objects is not as trivial as slapping basis vectors onto the outside of a set of  tensor components will become evident below.

Besides false statements about spinors and coordinate generality, one sometimes finds discussions that start out true and illuminating, become misleading (such as by being true for the wrong reason), and conclude with a false gloss.  The 1958 discussion by Gel'fand, Minlos and Shapiro is illustrative. In light of Ogievetsky-Polubarinov spinors,  I will make a running commentary in square brackets in italic font.
\begin{quote}
We note that the tensor representation of the Lorentz group\ldots may be extended to a representation of the entire group of non-degenerate linear transformations $a$ in four-dimensional space.  In order to do this it suffices to substitute into this formula the matrix of the linear transformation $a.$ \emph{[They seem to envision a rigid rather than position-dependent linear transformation, but that does not matter for present purposes; we are not yet worried about covariant and Lie derivatives.]} Thus, the tensors\ldots which we considered above only in orthogonal coordinate systems (i.e. those systems which transform into one another by Lorentz transformations) may be written down in any coordinate system.

The situation is different for the spinor representations\ldots, which are not equivalent to tensor representations\ldots.  These representations of the Lorentz group cannot be extend to a representation of the entire group of linear transformations of four-dimensional space. \emph{[This statement is somewhat vague; its truth hangs by a thread.  As will appear below, it is true for metrics with indefinite signature, not for the reason that they envisioned, but because one cannot take a suitable square root of a matrix with negative eigenvalues.  It is false for spaces with definite signature, if one introduces the metric (or its conformal part) as an additional field on which the spinor transformation may depend, and lets the transformation depend nonlinearly on the metric.  If one aims to ``extend'' a representation, why not extend it using what is needed to make it actually work?]}  Thus spinors which we defined only in an orthogonal coordinate system cannot be defined in a natural fashion in an oblique coordinate system.  \emph{[It isn't clear what counts as ``natural'' here, but oblique coordinate systems are indeed admissible, as long as a certain matrix does not acquire negative eigenvalues.  The resulting transformation law is linear in the spinor and nonlinear in the conformal part of the metric.]}  
\cite[p. 252]{GelfandLorentzSpinor}
\end{quote}
This work appeared before everything about Ogievetsky-Polubarinov spinors except the DeWitts' precursor \cite{DeWittSpinor}.

One can sometimes detect an undercurrent of dissatisfaction with the standard orthonormal tetrad formalism even among its advocates.  Lawson and Michelsohn observe that  
\begin{quote}
\emph{the bundle of spinors itself depends in an essential way on the choice of riemannian} [\emph{sic}] \emph{structure on the manifold.}  \\
 These observations lead one to suspect that there must exist a local spinor calculus, like the tensor calculus, which should be an important component of local riemannian [\emph{sic}] geometry.  A satisfactory formalism of this type has not yet been developed. \cite[p. 5, emphasis in the original]{SpinGeometry} \end{quote}
The dependence on the metric leads one not to be surprised by the way that Ogievetsky-Polubarinov spinors form a geometric object (up to a sign) only with the help of the metric (or rather, its conformal part, as will appear below).   Like many authors, Lawson and Michelsohn give no indication of awareness of Ogievetsky-Polubarinov spinors.  It is not difficult to conjecture that this formalism fills the hole sensed by Lawson and Michelsohn.  

Representing spinors with respect to coordinates in a curved space-time, which supposedly was shown impossible in 1929, is precisely what Ogievetsky and Polubarinov did in 1965.   
 The key technical assumption to be rejected by Ogievetsky and Polubarinov was that the spinor transformation law should depend only on the spinor, not also on the metric (or its conformal part, as it happens) \cite{OPspinorReprint}. 
It is quite evident that Cartan had in mind linear coordinate transformations for the spinor by itself \cite{CartanSpinor}. Once one gives up that assumption, which appears to be introduced implicitly due to understandable lack of imagination, rather than explicitly for any defensible reason, one finds that spinors almost fit do into tensor calculus.  In particular, they fit well enough that they have spinorial Lie and covariant derivatives due to their coordinate transformation properties \cite{OPspinorReprint}.  Making comparison of their spinor formalism to the  1950s-60s literature on geometric objects, including nonlinear geometric objects (in which the components in one coordinate system depend nonlinearly on those in another) \cite{Tashiro1,Nijenhuis,Tashiro2,Yano,AczelGolab,SzybiakCovariant,SzybiakLie}, explains various features of the Ogievetsky-Polubarinov (henceforth ``OP'') formalism.  The literature on nonlinear geometric objects, being motivated more by a quest for taxonomic completeness than by recognition of interesting examples, fell into neglect with the modernization of geometrical language.  Ironically, parallel ideas grew up about the same time among particle physicists, partly out of the original OP paper, and have flourished under the name of nonlinear group representations, where they appear in current  gauge theories of gravity.  

 OP's  constructive proof that finite-component spinors can be represented in coordinates in curved space-time has made rather less of an impact than one might have expected.  One finds to this day many eminent authors who repeat the Weyl-Cartan claim that such a thing is impossible 
 \cite{Weinberg,DeserVierbein,vanNReports,FaddeevEnergy,SpinGeometry,FronsdalMass1,Kaku,BardeenZumino,WeinbergQFT3,FatibeneFrancaviglia,FerrarisConserved}.  
What one evidently never finds is any critique of the OP formalism; clearly the same lack of imagination is still doing much of the work as did it in the 1930s, albeit less understandably. While the necessity of the tetrad formalism is still the dominant view, one cannot speak of any debate, but rather a mere failure to engage with the relevant literature by the majority and a surprising lack of polemic from those acquainted with OP spinors, like ships passing in the night.  But 46 years is long enough for the neglect of a constructive refutation of a famous no-go theorem.


A more correct picture  is provided by some workers in supergravity (a field where Ogievetsky and Polubarinov are well known \cite{OPspinorReprint}):
\begin{quote}
in fact it is impossible to define a field, transforming \emph{linearly} under the $\lambda$ [coordinate transformation] group, which also transforms as a spinor when the $\lambda$'s are restricted to represent global Lorentz transformations.  It is possible to get around this difficulty by realizing the $\lambda$ transformations nonlinearly, but this is not a convenient solution.  \cite[p. 234]{GatesGrisaruRocekSiegel}, emphasis in the original. 
\end{quote}
The reader interested in practical calculations, on contemplating the OP formalism to which Gates \emph{et al.} allude, might well doubt its convenience.  On the other hand, some worked examples below indicate that practical calculations actually need not be inconvenient, as it turns out, because the significant symmetries present in most problems addressed with pencil and paper make the formalism simplify. But more to the point, for \emph{ philosophical} analysis, it does not matter whether a formalism is convenient for explicit calculations by having a pleasing linear form, or instead takes a more demanding nonlinear form.  The philosopher pays special attention to conceptual problems \cite{LaudanProgress}, an area where the OP formalism has significant advantages and some surprises. 
Gates \emph{et al.} continue a bit later:
\begin{quote} To solve these problems we enlarge the gauge group by adjoining to the $\lambda$ transformations a group of \emph{local} Lorentz transformations, and define spinors with respect to this group.  This is a procedure familiar in treatments of nonlinear $\sigma$ models.  Nonlinear realizations of a group are replaced by linear representations of an enlarged (gauge) group.  The nonlinearities reappear only when a definite gauge choice is made.  Similarly here, by enlarging the gauge group, we obtain linear spinor representations.  The nonlinear spinor representations of the general coordinate group reappear only if we fix a gauge for the local Lorentz transformations.     \cite[p. 234]{GatesGrisaruRocekSiegel} \end{quote}
A bit later a remark relevant to the relationship between the OP formalism and the more common tetrad formalism appears:
\begin{quote}  It is thus possible to gauge away the antisymmetric-tensor part of the vierbein (although not the scalar part) with a \emph{nonderivative} transformation.  To stay in this [symmetric] gauge a local coordinate transformation must be accompanied by a related local Lorentz transformation; the Lorentz parameter is determined in terms of the translation parameter.  \cite[p. 235]{GatesGrisaruRocekSiegel} \end{quote}  The tetrad formalism provides a back door to the OP formalism, locally, by imposing the local Lorentz (that is, $O(1,3)$)-fixing condition that the components of the tetrad form a symmetric matrix, once an index is moved with the matrix $diag(-1,1,1,1)$ (or its opposite). \emph{Having spinors in coordinates is not impossible; it's just difficult and unfamiliar.}

 Mention of the DeWitts as providing a useful clue in the 1950s foreshadows a common theme, namely, ambiguity and implicit self-contradiction.   Many authors, apparently unaware of the OP formalism in most (not all) cases, have imposed the symmetric gauge condition on the tetrad  \cite{IshamSalamStrathdee,IshamSalamStrathdee2,DeserNonrenorm,tHooftVeltmanOneLoopAIHP,BorisovOgievetskii,ChoFreund,HamamotoNonlinear,HamamotoVector,DeserFermion,Veltman,IvanovNiederle,AlvarezGaumeWitten,Passarino,FujikawaAnomalies,FujikawaAnomalies,ChoiShimSong,AldrovandiNovaes,LPTT,Schucker,TresguerresMielke,BilyalovSpinors,Sardanashvily,ObukhovPereira,VasilievSymmetric,TiembloTresguerres2,Kirsch,Holstein,Leclerc,Iochum,Nibbelink}. Those who were unaware of the OP spinor formalism and imposed the symetric gauge on the tetrad  thus in effect partially reinvented the OP formalism.
 It is often not difficult to find the same author both  asserting that having spinors in coordinates is impossible and  implicitly showing (in the same paper or elsewhere) how to do it---by imposing a symmetric gauge condition on the tetrad
 \cite{DeWittDToGaF,DeserNonrenorm,DeserFermion,DeserVierbein,IshamSalamStrathdee,AldrovandiNovaes,Leclerc,LeclercNoether}.  Bryce DeWitt, having provided (with C\'{e}cile) perhaps the first substantial hint toward nonlinear metric-dependent spinors  \cite{DeWittSpinor}, later asserted the necessity of a tetrad \cite{DeWittDToGaF};  still later he retreated, in view of their own earlier work, to the   convenience claim later made by Gates \emph{et al.}, that  a nonlinear spinor transformation law is possible but not as convenient as a tetrad \cite{DeWitt67c}. What is perhaps most remarkable in this spectacle of contradictory claims passing like ships in the night is how rarely the contradictory views are simultaneously noticed and recognized as contradictory, a phenomenon that this paper aims to overcome.
 Clearly the authors who deny the existence of spinors in coordinates, while in effect showing how to have them, typically have not realized that they were sketching a formalism that is in principle independent of the tetrad formalism with which they started.  In effect many authors have largely reinvented the OP coordinate spinor formalism, even while denying  its possibility.  Such a phenomenon cries out for attention from historians and philosophers of science.  Thus far such attention has not been given, apart from the promissory note in \cite{FriedmanJones}, where OP spinors are used to escape a counterexample to the Anderson-Friedman absolute objects analysis of substantive general covariance \cite{Anderson,TLL,LLN,FriedmanFoundations}.  Histories of spinors in General Relativity still read as if the theory took its definitive form around 1930  \cite{KichenassamySpinor,ScholzWeylFockSpinor}.  On a somewhat related note to mine,  Ne'eman and Sijacki discuss the persistence of erroneous non-existence claims for spinors under the double covering of the general linear group \cite{NeemanSijacki,Sijacki}. While they are more interested in infinite-component linear spinors than finite-component nonlinear (in the metric) OP spinors, they make use of the latter as well. 
Part of the dispute, presumably, is verbal, depending on what counts as coupling directly to the metric (whether the concomitant $r_{\mu\nu}$ is `direct' enough and whether the author even knows about it or has entertained the idea that it might be a good object in its own right), what counts as a tetrad (whether $r_{\mu\nu}$, being symmetric and  numerically equal to a gauge-fixed tetrad almost everywhere on the manifold, is `just' a tetrad), 
what counts as a  representation (\emph{vs.} representation up to a sign, as one often sees with spinors), \emph{etc.}

This paper aims to make OP spinors well known.  It, moreover, aims to make them intelligible.  The causal reader who takes up the original OP papers, for example, might well be terminally perplexed; the persistent reader is richly rewarded. Apparently nothing has yet been written on OP spinors  that is relatively easy to understand, well connected to traditional tensor calculus concepts, and largely complete. This paper aims to fill that gap at least in outline.  The precise details, involving specific coefficients of some infinite series, will not be considered, however.  Instead some conceptual features that are by no means obvious on confronting the explicit mathematical form will be emphasized.  
In the original OP paper, besides the once-familiar imaginary time coordinate  $ x^4=ict$ that takes away the comforting pairing of contravariant and covariant indices, one finds perturbative expansions that particle physicists are more likely to countenance than are general relativists (and that implicitly privilege approximately Cartesian coordinates over other coordinates),   matrices that look vaguely like a Lie derivative multiplied by unfamiliar expressions,
a gravitational potential obtained by subtracting a constant numerical matrix from the components of the metric tensor (on which more below), and, in general, the appearance of strong coordinate dependence in a way that makes covariance anything but manifest.  Explanatory commentary is somewhat brief, while there is no  connection drawn to the nonlinear geometric objects literature, which might have been invoked to provide a natural home for OP spinors within ongoing research in differential geometry in the mid-1960s.  Neither do OP show  any interest in coordinate transformations beyond infinitesimal order, this being a genuine omission, as opposed to the above list of primarily pedagogical challenges. There is little interest in the convergence of the series expansions, either (\emph{c.f.} \cite{DeWittSpinor}), another substantive question leaving one to wonder whether, for example, polar coordinates are admissible.  (They are, as becomes clear with Bilyalov's generalized eigenvalue formalism.  In fact, if the metric's signature were positive definite, all coordinates would be admissible, as usual in tensor calculus.)
 But the reader who perseveres and ponders the formalism is rewarded with deep conceptual insights as well as interesting open problems.  

One problem left open by OP has been taken up by Ranat Bilyalov, namely, finite and even large coordinate transformations \cite{BilyalovConservation,BilyalovSpinors}.  He has shown that, while any coordinate transformation is admissible near the identity, not just any is admissible in the finite case, because not  just any coordinate system is admissible.  Recall that coordinates are always described as ordered $n$-tuples.  Generally no importance is attached to the order in the presence of a curved metric (or even a flat metric in non-Cartesian coordinates), but there is an order nonetheless.  It turns out in the OP formalism as elaborated by Bilyalov, due to nonlinearity and the role of the signature matrix $diag(-1,1,1,1)$ that treats the $0th$ coordinate differently with the minus sign (here used as an alternative to $x^4 = ict$), that not just any coordinate can be first; sometimes one needs to swap the coordinates, as will be discussed in  detail below.  Thus electrons, protons and the like care about `time' in a way that photons and gravitons do not.  There is a great deal of flexibility about what can be a time coordinate, but not total flexibility as one might expect from the precedent of tensor calculus or the usual (de-Ockhamized) tetrad spinor formalism.

A transition from the OP spinor formalism to the tetrad formalism would illustrate part of the conventional wisdom often associated with Kretschmann \cite{Norton}, namely, that arbitrary coordinates are admissible in any theory with some ingenuity. 
 In this case one can have the time coordinate in what used to be the `wrong' place by introducing 6 extra components to build a tetrad, and letting spinors be spinors with respect to a new gauge group unrelated to space-time. The new spinors are coordinate scalars and so do not care about the order of the coordinates.

The other part of the conventional wisdom often linked to the name of Kretschmann holds that  admissibility of any coordinates into any theory supposedly is trivial; one needs only to use tensor calculus (\emph{e.g.}, \cite{EarmanCovariance}).  That last claim, that it's easy because only tensor calculus is needed, is obviously false if one reflects on spinors while thinking about general covariance.  Having to introduce a tetrad to fill an inadequacy of the tensor calculus, as Weyl  urges as a theorem, is hardly an instance of easily solving a problem using tensor calculus.  But reflection on spinors rarely happens in the context of discussions of formal \emph{vs.} substantive general covariance, largely because particle physics is wrongly thought irrelevant to space-time theory and most general relativists do not write a lot about spinors.\footnote{In view of the fact that in quantum field theory, spinors generally are almost anti-commutative rather than almost commutative, ideally one should consider anticommuting `classical' spinors.  But I will not worry about that additional wrinkle.}

 What of the claim that the admissibility of any coordinates is trivial? It is not perfectly clear whether  general covariance is a substantive physical claim about a space-time theory, or merely a formal feature that any theory, suitably formulated, can have.  Thus some authors
 \cite{BergmannLectures,Anderson,FriedmanFoundations,StachelGC,NortonGC,EarmanCovariance,PittsArtificial} 
have contemplated criteria for distinguishing the substantive general covariance that supposedly distinguishes Einstein's General Relativity (GR) from pre-GR theories, from the merely formal general covariance that any theory presumably can have.

 Here one needs to consider what price is to be paid for such admissibility of any coordinates.  One can draw an analogy to internal gauge symmetries, such as in electromagnetism.  One can install various gauge symmetries in theories that naturally lack them, simply by introducing gauge compensation fields---cheating, one might say.  If the gauge compensation fields zag when the physical fields zig, so to speak, then one has a theory with gauge freedom that is physically equivalent to the original.  The most famous example is the Stueckelberg trick in massive electromagnetism   \cite{Ruegg,PittsArtificial}.  Massive electromagnetism, usually associated with Proca, adds a term $-\frac{m^2}{2} A^{\mu} A_{\mu}$ to the Lagrangian density $F^2$ for Maxwell's electromagnetism, with  $m$ being the ``photon mass.''  The mass term breaks the gauge symmetry.  That gauge freedom can be restored through the introduction of a  gauge compensation field, a gauge-dependent coordinate scalar. The Stueckelberg mass term takes the form  $ -\frac{m^2}{2} (A^{\mu}  - \partial^{\mu} \psi) (A_{\mu}  - \partial_{\mu} \psi)$. A gauge transformation of $A_{\mu}$ is compensated by changing $\psi$:  $A_{\mu} \rightarrow A_{\mu} + \partial_{\mu} \chi ,$ $\psi \rightarrow \psi + \chi.$   Once again it is possible to admit `all' descriptions on equal footing, but the artifice of the gauge compensation field $\psi$ is palpable. (One notices that $\psi$'s Euler-Lagrange equation is redundant and that one can choose the gauge $\psi=0$ without spoiling locality or Lorentz invariance, for example.) Were someone to suggest that the availability of the Stueckelberg trick shows that electromagnetic gauge symmetry is trivial, because even Proca's theory can be reformulated so as to have such gauge symmetry, the palpable artificiality of the gauge compensation field $ \psi$ shows otherwise.   To use such fields is to cheat at gauge invariance; the difference between the natural gauge symmetry of Maxwell's electromagnetism and the artificial gauge symmetry of massive electromagnetism with Stueckelberg's trick is evident.  One might propose  that having coordinate or gauge symmetry without cheating \emph{via} gauge compensation fields should be considered the substantive notion of general covariance or gauge symmetry \cite{PittsArtificial}.

While it is true that one can have general coordinate symmetry or some other gauge symmetry into `any' theory, suitably formulated---I won't attempt to specify just what theories are being quantified over, but local (not necessarily relativistic) field theories would be a good proposal---the cost of \emph{installing} it can be high.  The theory might \emph{want} to teach a contrary lesson, as in massive Proca electromagnetism;  it would be perverse to insist on using the Stueckelberg formulation on the basis of a supposed generalized Kretschmann lesson about gauge freedom.  By contrast, it is not perverse, but quite reasonable (and common), to employ the Stueckelberg formulation for certain calculations in quantum field theory, such as establishing renormalizability of massive electromagnetism. High energy physicists have deployed considerable ingenuity in developing general formalisms that install artificial gauge freedom, such as  the Batalin-Fradkin-Tyutin formalism.  The de-Ockhamized formalism can be very useful technically, but is quite unhelpful conceptually.


\subsection{Tetrad in Symmetric Gauge as Back Door to OP Spinors}

It is somewhat curious that the widely accepted no-go claim about having spinors see a curved metric without a tetrad and the explicit refutation by construction in 1965 have managed to coexist so peacefully for so long. In fact one can find hints of the OP formalism even earlier \cite{DeWittSpinor,Huggins}.  A number of authors have unwittingly largely reinvented the OP formalism by imposing the symmetric gauge on the tetrad formalism:  $e^{\mu A} =  e^{\alpha M}$, with the indices moved with  $\eta$.  (My notation here employs an obvious Greco-Roman alphabet correspondence principle.  Some authors who impose this gauge condition prefer to say that two types of indices have merged.  Often  the OP formalism requires doing things that standard tensorial notations are designed to prevent.)  This fixes the local $O(1,3)$ gauge freedom.

The usual tetrad formalism involves both coordinate transformations and local Lorentz transformations, with the resulting mixed transformation rule 
\begin{equation}e^{\mu^{\prime} }_A = \frac{\partial x^{\mu^{\prime} } }{\partial x^{\nu} } e^{\nu}_B J^B_A,
\end{equation}
where $J^B_A \eta_{BC} J^C_D = \eta_{AB}.$
$J^B_A$ is an arbitrary position-dependent element of $O(1,3)$. 

 To approach the OP formalism, one needs to choose $J^B_A$ in each coordinate system such that the tetrad matrix is symmetric, once the Latin index is moved with $diag(-1,1,1,1).$ The resulting transformation rule for the symmetrized tetrad is nonlinear: it has a  linear factor because of the vector index of the tetrad, but acquires further dependence on the tetrad \emph{via} $ J^C_D,$ which is chosen to yield a symmetric matrix in the given coordinate system.  Gauge-fixing the local Lorentz freedom in a coordinate- and tetrad-dependent way yields a symmetrized tetrad with a  nonlinear transformation law, which is basically the symmetric square root of the inverse metric $r^{\mu\nu}.$  It is easy to do so explicitly in $1+1$ dimensions, where there are no rotations and the boosts all commute; remarkably, it seems to be possible to say something explicit in closed form more generally \cite{LPTT}.

One then notices that there is no need to introduce the tetrad in the first place; why not simply deal with the symmetric quantity instead?  That symmetric entity $r^{\mu\nu}$ or $r_{\mu\nu}$ \emph{itself} represents arbitrary infinitesimal coordinate transformations, nonlinearly, with only coordinate indices.  At some stage one should consider just how generally the symmetric gauge can be reached from the tetrad; the answer will determine the admissible coordinates in the OP formalism. 

 The fact that an orthonormal basis can be topologically obstructed on manifolds with a metric shows that the symmetric square root of the metric is conceptually independent of a tetrad.  As simple a case as the 2-sphere with positive definite metric makes the point, with the hairy ball theorem excluding even one nowhere-vanishing vector field, to say nothing of a basis.  The symmetric square root of the metric is an analytic function of the metric, as shown by the perturbative expansion, and thus exists (patched together as needed from chart to chart) globally wherever the metric does.  Clearly the symmetric square root of the metric is far from being `just' a gauge-fixed tetrad, because the former exists even in cases where a tetrad does not. 

\subsection{Generalized Polar Decomposition}

Componentwise, one can obtain the symmetric square root of the metric using a generalized polar decomposition of the tetrad, treated as a matrix of components.   The usual polar decomposition factors a positive definite matrix into an orthogonal (rotation) factor and a symmetric (shear and expansion) factors.  (Clearly the expansion, which corresponds to the volume element in our applications, goes along for the ride and can be separated out using a unimodular matrix and a scalar factor. Below it will be noted that it cancels out altogether from the Dirac equation for massless spinors.) 
Isham, Salam and Strathdee have invoked the polar decomposition near the identity, but without mathematical control over what happens for large transformations \cite{IshamSalamStrathdee,IshamSalamStrathdee2}.  Progress in applied mathematics and linear algebra has largely filled that hole.  

A generalized polar decomposition makes use of a ``signature matrix'' like $\eta=diag(-1,1,1,1)$ \cite{Bolshakov1,Higham}. A matrix $M$ is $\eta$-orthogonal iff $M^T \eta M = \eta$.  It is called $\eta$-symmetric if symmetric with index moved by $\eta$. 
 Tweaking some  notation a bit, one has the following theorem applicable to generalized polar decomposition of a tetrad into a symmetric square root and a boost-rotation: 
\begin{quote} Theorem 5.1. If [tetrad component matrix] $E \in  R^{n \times n}$ and $\eta E^{T}\eta E$ has no eigenvalues on the nonpositive real axis, then $E$ has a unique indefinite polar decomposition $E = QS$, where $Q$ is
$\eta$-orthogonal and $S$ is $\eta$-symmetric with eigenvalues in the open right half-plane. \cite[p. 513]{Higham} 
\end{quote}
One can find the boost-rotation explicitly:   $Q=E(\eta E^T \eta E)^{-\frac{1}{2}}$ fixes  the local $O(1,3)$ transformation.
Having made this decomposition of the tetrad $E,$ the OP formalism works with $S$ or $R = \eta S,$ throwing away the local $O(1,3)$ factor, which is pure gauge, and keeping the symmetric factor, which is the symmetric square root of the metric, a physical quantity equivalent to the metric.

\subsection{Identifying and Eliminating Surplus Structure}

One common theme in modern physics is the value of identifying and eliminating surplus structure.  One class of examples that has come up in both Newtonian and relativistic contexts is the elimination of a preferred reference frame with no observable consequences.  Newton's absolute space and the aether rest frame are textbook examples of entities that, though perceived by their advocates to have some explanatory advantages, have eventually been regarded as superfluous and worthy of elimination.  James Anderson's absolute objects analysis of general covariance, for example, has emphasized the importance of eliminating irrelevant fields in order to identity absolute objects, find the symmetry group, \emph{etc.} \cite{Anderson,TLL,LLN,FriedmanJones}.

This work takes notice of two distinct compatible ways of eliminating surplus structure from the typical formulation of the massless Dirac equation in a curved space-time.  One entity that can be eliminated, as was realized by Haantjes and Schouten in the 1930s \cite{SchoutenHaantjesConformal,HaantjesConformalSpinor} but apparently forgotten afterwards in the move to modern notation, is the volume element $\sqrt{-g}.$  Many authors nowadays discuss the \emph{co}variance of the Dirac operator (the left side of the massless Dirac equation) under conformal changes of metric, but (almost?) no one points out that one could, using a suitable choice of primitive fields, achieve \emph{in}variance, in which $\sqrt{-g}$ disappears altogether from the theory.  This difference is important for two reasons.  First, it is widely believed that a shift from covariance under arbitrary coordinate transformations to invariant objects in modern differential geometry was a great leap forward. Why introduce surplus structure only to show its irrelevance? The general failure to treat the volume element that way implies that either the shift from covariance to invariance was not a great leap forward, or that modern formulations of the Dirac equation are inferior to some from  the 1930s in important respects; either answer suggests a reform of current practice.  Second, frequently one finds authors drawing conclusions showing that they do not infer from \emph{co}variance under conformal rescaling of the metric that the volume element is irrelevant.  That is evident whenever someone assigns a special status to Killing vector fields (fields such that the Lie derivative of the metric tensor vanishes) that is not assigned to conformal vector fields (fields such that the Lie derivative of the conformal part of the metric, which precisely fixes the null cone, vanishes)---which does happen  \cite{KosmannLie,KosmannLieApplied,FatibeneFermion,Cotaescu,FatibeneFrancaviglia}.  Given the eliminability of the volume element from the massless Dirac equation, one no more needs cooperation from $\sqrt{-g}$ (as in Killing's equation) than one needs cooperation from the aether rest frame in special relativistic physics.  Suitably purged of irrelevant fields, the massless Dirac equation does not know (locally!) whether it is in a flat space-time or a merely conformally flat one.

There is an interesting difference between the two purgings of surplus structure from the massless Dirac equation.  Neglecting gravitation, the volume element is not varied in the principle of least action.  Eliminating it therefore eliminates a non-variational field, which thus increases the symmetry group of the theory from the symmetries of the metric to the symmetries of the conformal part of the metric.  In flat four-dimensional space-time, that is an increase from 10 Killing vector fields to 15 conformal Killing vector fields.  
The antisymmetric part of the tetrad, by contrast, serves as a gauge compensation field, much like the Stueckelberg scalar $\psi$ that installs artificial electromagnetic gauge freedom. Eliminating the antisymmetric part of the tetrad (as in the OP formalism) thus removes the local Lorentz ($O(1,3)$) symmetry.  Removing $\sqrt{-g}$ makes a real symmetry manifest, while symmetrizing the tetrad removes an artificial symmetry.  

%

\subsection{General Covariance and Coordinate Reordering}

General coordinate transformations contain a certain collection of transformations that probably  no one has ever regarded as interesting.  They got into the formalism by accident.  Indeed perhaps no one has noticed them prior to Ranat Bilyalov's work on spinors \cite{BilyalovConservation}, and few have done so afterwards.  That collection involves simply \emph{reordering} the coordinates.  
The usual formalisms require that the coordinates be expressed in some definite order, but no significance is attached to that order.  For convenience  we can focus attention on Special Relativity for the moment; similar lessons apply in General Relativity, in many respects. Coordinates are listed as an \emph{ordered} quadruple.  
If one calls the original coordinates $(t, x, y, z)$ for example, with $t$ being the standard temporal coordinate in an inertial frame and $x,$ $y,$ and $z$ being spatial Cartesian coordinates as the notation suggests,  then the new coordinates after reordering might be $(t, x, z, y).$  In this case the reordering of two spatial coordinates has occurred; this change is so trivial that one need not make any change to the metrical matrix $diag(-1,1,1,1)$ (with the speed of light set to $1$) to accommodate it. 

 Slightly less trivial  is a reordering that involves the time coordinate:  $(t, x, y, z)$ might be replaced by $(x, t, y, z).$ Now the  matrix $diag(-1,1,1,1)$ is inappropriate; the invariant interval in the new coordinates is given by $diag(1,-1,1,1).$   In short, one needs tensor calculus, or a small fragment  of it, namely the part that introduces a metric tensor and its coordinate transformation law, in order to permit the coordinate reordering.  One does not need a connection; the Christoffel symbols in the new coordinates are still $0$ because the new coordinates are linear functions of the old.  One certainly could formulate Special Relativity in this fashion.  Presumably no one would bother to introduce a metric tensor simply for the slight gain in generality of admitting reorderings of the Cartesian coordinates, however.  The indefinite signature of the metric, which distinguishes temporal from spatial coordinates, is crucial here; no analogous issue arises for positive definite metrics.

I labor the modest point about coordinate reorderings  because the freedom to write `time' (suitably generalized) as the second, third or fourth coordinate (in four space-time dimensions) is  what one \emph{gives up} in the Ogievetsky-Polubarinov (OP) spinor formalism 
 \cite{OPspinorReprint,BorisovOgievetskii,BilyalovSpinors}.   It seems to me that no one should miss that rather trivial bit of coordinate freedom. So if such freedom is excluded by a procedure that is very beneficial overall, the benefit greatly outweighs the cost.  If that  proposal is accepted, then in the presence of spinors one should renounce talking about coordinate transformations involving the general linear group $GL(4, R),$ whether rigid or position-dependent, in favor of some collection that divides out by the coordinate reorderings  involving time. The  condition for coordinate admissibility, as will appear below, thus depends on the value of the metric components, assuming indefinite signature.  The result  will most likely be a Brandt groupoid,  a structure that would be a group except that not every pair of elements can be multiplied \cite{Hahn,Renault}.  
 Of course the infinitesimal transformations and associated Lie algebra will not notice the difference; reordering the coordinates is not a change that one builds up from infinitesimal pieces.  On the other hand, if one wishes to preserve the usual sense of general covariance as admitting arbitrary coordinate transformations, then this modest freedom to reorder the coordinates is the increase.  
It  is evident that, contrary to the usual practice of admitting all possible coordinates \cite{Wald}, it is fitting to take the atlas of admissible coordinate charts to vary with the types of fields present:  in the presence of spinors and an indefinite metric, one should throw out coordinate interchanges involving time, whereas in the absence of spinors or with a positive definite metric, such transformations are harmless.

When one contemplates position-dependent coordinate transformations, as one may in SR and must in GR, the requirement to keep `time' first will be somewhat more demanding, in that charts such that the `time'-like character perhaps passes back and forth between different coordinates will be excluded---though pieces without such fluctuation will still be admissible. It follows that one does not know whether a coordinate system is physically admissible until the dynamical problem is solved.  One can admit all coordinates provisionally, but some prove \emph{a posteriori} not to reflect the underlying physics adequately.


That arbitrary coordinates are admissible in GR is not merely conventional wisdom---it is one of the theory's most striking and famous features.  Yet I argue that this is a mistake, albeit a small one, in the presence of spinor fields.  Of course fermions such as electrons, protons and neutrons are hardly negligible or  unusual things, notwithstanding their scarcity in philosophers' discussions of space-time theory.  So any adequate understanding of space-time theory must take spinor fields into account; the unfortunate neglect of particle physics has helped to keep this important issue from attracting the attention that it deserves.  

To understand the good and the bad considerations involved in the usual and by now largely uncontroversial admission of arbitrary coordinates, it is helpful to remember an argument that Einstein made in 1916 when the issue was novel: 
\begin{quote}  The method hitherto employed for laying co-ordinates into the space-time continuum in a definite manner [yielding observable time or space intervals] thus breaks down, and there seems to be no other way which would allow us to adapt systems of co-ordinates to the four-dimensional universe so that we might expect from their application a particularly {\bf simple} formulation of the laws of nature. So there is nothing for it but to regard all imaginable systems of co-ordinates, on principle, as equally suitable for the description of nature. 
\cite[p. 117, emphasis added]{EinsteinFoundation} \end{quote}
Indeed coordinates with \emph{quantitative} physical meaning are not available with adequate generality in the curved space-times of GR.  The next crucial step involves inferring the nonexistence of something, given the failure thus far (in 1916) to imagine it.  This step was eminently reasonable in 1916 and indeed for some time afterward; but clearly the argument is defeasible.
There could conceivably arise some \emph{qualitative} physical meaning that coordinates retain, or usefully may retain, even in General Relativity.  
 Note also that the amount of coordinate freedom, one might say the scope of conventionality, is itself conventional:  having admitted sufficiently general coordinates to cover curved space-times, one decides whether or not to impose some restrictions based on whether simplicity can be achieved thereby.  Of course in SR the quest for simplicity is naturally realized using Cartesian coordinates.  In GR there seemed to be no analogous simplification available, so admitting arbitrary coordinates was the natural conclusion.  However, by now, plausibly this impossibility of imagining a formulation of the laws of nature that is simpler with somewhat restricted coordinates then with arbitrary coordinates has in fact been overcome by mathematical developments after mid-century.

  OP spinors, though not always simple in terms of the process of explicit calculation, are simple in the sense of substantially unifying spinors with tensors and the like and in the sense of allowing spinors to be treated without introducing six extra field components and then six extra symmetries to gauge them away.  OP spinors simply exclude surplus structure, rather than introducing it and then marking it as surplus with an additional gauge group. That, surely, is conceptual progress.   Whereas the tetrad spinor formalism makes spinors behave spinorially in terms of a non-coordinate index on the tetrad $e^{\mu}_A$ (here written as a large Latin index rather than a small Greek index), the OP formalism makes spinors behave spinorially with respect to coordinate transformations.

\section{Geometric Objects, Especially Nonlinear Ones}

\subsection{Geometric Objects}

The development of tensor calculus led eventually to the concept of a geometric object \cite{Nijenhuis,Schouten,AczelGolab,Trautman,Anderson,Golab}, which received substantially satisfactory definition during 1934-7 \cite{SchoutenHaantjes}.  The tensor calculus, which Weyl and Cartan had proclaimed unable to accommodate spinors, had not reached its natural limits when they wrote.  
The  theory of geometric objects, which was largely complete in the 1960s in the linear case, 
describes tensors and more general geometric objects, including connections, as well as some little known nonlinear entities, some but not all of which are equivalent to linear ones.  
This theory, which was never well known, has been partially forgotten in the wake of the modern style, except for a few papers on ``natural bundles,'' such as  \cite{NijenhuisYano,FatibeneFrancaviglia}, which are far from exhaustive in applications, so some review will be appropriate.

The following is a typical definition of a geometric object (neglecting some finer distinctions \cite{Nijenhuis} and qualifications \cite{KucharzewskiKuczma}): it is, for each space-time point and each local coordinate system around it, a finite ordered set of   components and a transformation law relating components in different coordinate systems at the same point \cite{Nijenhuis,KucharzewskiKuczma,Trautman}.  (In more modern terms, it is a global, as opposed to local, section of an appropriate bundle, a fact not always helpful in dealing with empty spaces around fluids or elastic solids \cite{FriedmanJones}. One winds up having a four-velocity of nothingness, which presumably ought to be $0$ or arbitrary if it is defined at all.  I won't attempt to modify the definition to avoid this problem, however.)
For example, a tangent or ``contravariant'' vector field $v^{\mu}$ has the coordinate transformation law 
$$v^{\mu^\prime}(p) = \frac{ \partial x^{\mu^\prime} }{\partial x^{\nu}  }(p) v^{\nu}(p)$$
for obtaining the components in primed coordinates $ x^{\mu^\prime}$
in terms of unprimed coordinates $ x^{\nu}$ at point $p$. 
More generally, tensors have linear homogeneous transformation laws: $v^\prime \sim v,$ where the usual flurry of indices is suppressed, partly for generality.
A linear homogeneous transformation law also exists for  tensor densities \cite{Anderson}, for which the transformation law involves some power of the determinant of the matrix $\frac{ \partial x^{\mu^\prime} }{\partial x^{\nu}  } .$
Geometric objects, as defined here, are not intended in this tradition to represent everything that one might wish to talk about, but only everything that is `like a tensor field' in some initially somewhat inchoate sense.

If one wishes to take a covariant derivative, one does so using an affine connection, a geometric object with a linear but inhomogeneous (one might say ``affine'') transformation law   $\Gamma^\prime \sim \Gamma + O(0),$  where the term abbreviated as ``O(0)'' depends on the first and second partial derivatives of one coordinate system with respect to another (hence vanishing for linear coordinate transformations), but not the components $\Gamma.$  These are all the geometric objects that are in wide circulation.  

In what follows it will be helpful to introduce some slightly unusual notation.  There are contexts in which constants or constant matrices play a role in the theory of geometric objects, such as for certain inhomogeneous or nonlinear transformation laws.  The main quantity that will be needed here is the \emph{matrix} $diag(-1,1,1,1).$ (Whether there are interesting nonlinear geometric objects that do not involve something playing an analogous role is unclear.)  
At times this will be written as $\eta_{\mu\nu}$ or  $\eta^{\mu\nu}$,  if emphasizing its similarity to tensors is useful.  The key point to emphasize is that it is just a matrix, a collection of numbers, namely $1,$ $0,$ and $-1$;  it is not a metric tensor and has no transformation law, but  is always and everywhere the same.   One other occasions it will be helpful to emphasize how this matrix is \emph{unlike} a tensor, at which point using Latin indices will be helpful:  $\eta_{MN}$  or $\eta^{MN}.$  Especially when there are two kinds of indices, such as one finds in the tetrad formalism, writing $\eta$ with Latin indices will be helpful.  Somewhat awkwardly, the theory of nonlinear geometric objects, linear for a subgroup, involves cases where Greek coordinate indices and Latin indices merge \cite{OPspinorReprint,ChoFreund}. 
 In such contexts I will use  Greco-Roman alphabetic correspondence, as noted above, such that where there is an obvious correspondence between a Greek letter (such as $\alpha$ or $\mu$) and a Latin index (such $A$ or $M$, respectively---not worrying about capitalization), such indices can be added or contracted. 
The depths of geometric object theory require overcoming some of the  usual conventions of tensor calculus that are intended to assist the expression of coordinate covariance in more typical circumstances, in order to express coordinate covariance in more unusual circumstances. 


A simple but uncommon sort of geometric object is a tensor or tensor density less some constant or constant matrix.  For example, $\gamma_{\mu\nu} =_{def} g_{\mu\nu} -\eta_{\mu\nu},$  where $\eta_{\mu\nu} =diag(-1,1,1,1),$ can be used as the gravitational potential \cite{OP,OPspinorReprint}. One readily finds the transformation law
\begin{eqnarray}
\gamma^{\prime}_{\mu\nu} = \gamma_{\alpha\beta} \frac{ \partial x^{\alpha} }{\partial x^{\mu^{\prime} } } \frac{ \partial x^{\beta} }{\partial x^{\nu^{\prime} } }    + \eta_{\alpha\beta} \frac{ \partial x^{\alpha} }{\partial x^{\mu^{\prime} } } \frac{ \partial x^{\beta} }{\partial x^{\nu^{\prime} } }  -\eta_{\mu\nu}.
\end{eqnarray}
Another example, a close relative of which has been used in the Hamiltonian treatment of General Relativity \cite{DiracFixation}, is $k =_{def} \sqrt{h} -1,$
where $\sqrt{h}$ is the spatial volume element.  The resulting transformation rule is 
$$k^{\prime} = k \left| \frac{ \partial x}{\partial x^{\prime} } \right| + \left| \frac{ \partial x}{\partial x^{\prime} } \right| -1.$$ 
In both examples the point is to subtract away a background or `vacuum' value of a quantity, its perturbation from the constant value being of primary interest. One might want such an entity because, \emph{e.g.}, nonlinear stability theory involves a choice of variables that vanishes at the origin \cite{KhazinShnolBook}, in order that powers of the variables be good indicators of the magnitude of the terms.  For approximately Cartesian coordinates, the value of the perturbation behaves intuitively.  In all cases it behaves as a geometric object, inheriting convenient tensorial Lie and covariant differentiation properties, because such differentiations tend to shave off  terms of zeroth order in the fields from the transformation law \cite{Yano}. Thus these examples have Lie and covariant derivatives that are tensorial (a symmetric covariant tensor and a weight 1 density, respectively).  This example exhibits a moderately useful geometric object that exploits the features of the theory that go beyond the usual examples of tensors, maybe tensor densities, and connections.

 If one wishes to view geometric objects as `machines' that one feeds basis vectors into slots \cite{MTW}, then these machines involve adding a numerical piece with no slots and a piece with two slots in the case of $\gamma_{\mu\nu};$  for $k$ one has to add a numerical piece and a piece with four slots for a basis.  Densities with integral weights can be regarded as having still more slots, as powers of a top form or the like.  It appears to be meaningless to feed in a fractional or irrational number of basis vectors, however.  Densities with non-integral weights are useful.   Bryce DeWitt once found that his quantum gravity formalism simplified (in the sense of having only finitely many nonvanishing bare vertex functions in the absence of sources) using either $(-g)^\frac{5}{18} g^{\mu\nu} - \eta^{\mu\nu}$ or $(-g)^{-\frac{5}{22}} g_{\mu\nu} - \eta_{\mu\nu}$ as primitive fields \cite{DeWitt67b}. The simplest (and only probably extant) derivations of various classical theories of massive gravitons also use densities of irrational weights \cite{OP,MassiveGravity1,PittsScalar}.
Densities with nonintegral weights, if viewed as machines into which basis vectors are fed to give numbers as output, can and presumably must be defined as \emph{powers} of densities with integral weights; one is in effect raising an operator to a power, a procedure that is merely formal until one feeds the basis in. That is, one must feed in a basis, and only \emph{afterward} raise the resulting \emph{number} to a nonintegral power. 
Such an issue arises in defining ``half-forms'' in geometric quantization \cite{KostantHalfForm,Woodhouse}, a half-form being  a scalar density of weight $\frac{1}{2}.$  The virtue of half-forms  is that the product of two of them gives a covariant integrand suitable for integrating to get a scalar; if one wants to generalize Hamiltonian methods by giving up a polarization into $q$'s and $p$'s (of weight $w$ and $1-w,$ respectively, in the standard Hamiltonian  formalism), then one wants $w=\frac{1}{2}.$ 
The same problem of treating fractional weights arises when one wants to decompose a geometric object into its irreducible parts, once a standard exercise with the metric tensor \cite{Anderson}, where one finds a scalar density giving volumes and a tensor density $\hat{g}_{\mu\nu}$ defining the null cone.
 It is interesting (and convenient) that the coordinate transformation laws for densities do not make such distinctions between formal non-integral powers of operators for non-integral weights and more straightforward definitions for integral weights that a modern-style definition requires. Such bifurcation stands in contrast to tensors, whether covariant or contravariant, which can be regarded as boldfaced entities in their own right (without changing the subject to how they act on an arbitrary bit of surplus structure, surplus structure that might not even be globally defined, as nonvanishing vector fields are not on a 2-sphere, for example  \cite{Spivak1},  and orthonormal bases are not without Stiefel-Whitney class restrictions \cite{DeWittIsham}). Thus some of the elegance of the modern style is an artifact of unnatural restrictions on the entities employed.  If one considers the `other' kind of tensors---pseudoscalars, axial vectors, and the like---then one  evidently should  regard the `coordinate-free' entity as an  operator waiting to eat a basis and say whether it is positively or negatively oriented.  
Axial vectors cannot be ignored in particle physics and play a role in supergravity. 

%

\subsection{Nonlinear Geometric Objects}
Besides linear and affine geometric objects, there are  geometric objects with nonlinear transformation laws: the new components are \emph{nonlinear} functions of the old components, as well as depending on  $\frac{ \partial x^{\mu^\prime} }{\partial x^{\nu}  }(p)$ and the like.  Iterating the transformation rule, it is easy to see that polynomial laws are not an option, because the highest power will keep rising with the iteration.  Two remaining possibilities (perhaps among others \cite{AczelGolab,FunctionalEquations}) are fractional linear transformations and infinite series; one could perhaps also take the ratio of two infinite series.  Consider the symmetric square root  $r_{\mu\nu}$ of the metric, defined implicitly in any coordinate system (or any admissible one, a question requiring additional attention below) by  
\begin{equation} 
g_{\mu\nu} = r_{\mu\alpha} \eta^{\alpha\beta} r_{\beta\nu},
\end{equation}
  where $\eta^{\alpha\beta}=diag(-1,1,1,1)$ here is, as always in this work, only a matrix, not a metric tensor or field of any sort. 
This entity $r_{\mu\nu}$ exists at least  in many coordinate systems, most obviously in those not terribly far from freely falling Cartesian coordinates.  If one expresses the metric $g_{\mu\nu}$ as some perturbation about the matrix $\eta_{\alpha\beta}=diag(-1,1,1,1),$ then one can use the binomial theorem to find an infinite series expression for $r_{\mu\alpha}.$ The result (when the perturbation is small enough for the series to converge \cite{DeWittSpinor}) is \cite{OPspinorReprint}
\begin{eqnarray} r_{\mu\nu} =
 \sum_{k=0}^{\infty} \frac{ \frac{1}{2}! }{ (\frac{1}{2}-k)! k! } [(g_{\mu\bullet}-\eta_{\mu\bullet}) \eta^{\bullet\bullet}\ldots (g_{\bullet\nu}-\eta_{\bullet\nu})]^{k\ factors \ of \ g} \nonumber \\ = \eta_{\mu\nu} + \frac{1}{2}(g_{\mu\nu}-\eta_{\mu\nu}) - \frac{1}{8}(g_{\mu\alpha}-\eta_{\mu\alpha}) \eta^{\alpha\beta} (g_{\beta\nu}-\eta_{\beta\nu}) + \ldots. \label{Series} \end{eqnarray}
The factorial-like expression $\frac{1}{2}!/ (\frac{1}{2}-k)! $ can  be cashed out in terms of Gamma functions, but need not be because it stands for $\frac{1}{2} \cdot (\frac{1}{2} -1) \cdot \ldots \cdot(\frac{1}{2}-k + 1).$  If one seeks the coordinate transformation law for $r_{\mu\alpha},$ it follows from the metric transformation law for $g_{\mu\nu},$ which is linear and well known, and the definition as applied to both coordinate systems: 
$g=r \eta r$  and $g^{\prime} = r^{\prime} \eta r^{\prime}.$  There is no prime  on $\eta$ in the  expression for $g^{\prime}$, because $\eta_{AB}$ is the numerical matrix $diag(-1,1,1,1).$   Less schematically, one can write
$$r_{\mu\alpha}^{\prime} \eta^{\alpha\beta} r_{\beta\nu}^{\prime} = \frac{\partial x^{\alpha} }{ \partial x^{\mu\prime} } r_{\alpha\rho} \eta^{\rho\sigma} r_{\sigma\beta} \frac{ \partial x^{\beta} }{ \partial x^{\nu\prime} }.$$  
  This result is correct in general, but is still somewhat implicit in that $r_{\mu\alpha}^{\prime}$ does not appear alone on the left side. Note that the matrix $diag(-1,1,1,1)$ puts time first; if one tries to order the coordinates otherwise, then either one has to replace the matrix $diag(-1,1,1,1)$ with something else (an option that does not seem fruitful or economical), or one gets  perturbations of magnitude $\pm 2$ (perhaps disastrous even outside perturbative expansions) for no benefit even when there is no gravitational field present.  Note that Bilyalov presents a \emph{theorem} \cite{BilyalovConservation} regarding the necessity \emph{and sufficiency} of reordering the coordinates, not simply a plausibility argument for its necessity.  The coordinate reordering is in effect part of the service rendered by his  matrix 
\begin{eqnarray} T = 
\left[
\begin{array}{cccc}
0  & -1 & 0 & 0 \\ 
1 & 0 & 0 & 0 \\
0 & 0 & 1 & 0 \\
0 & 0 & 0 & 1 \end{array}
\right]_{,}  \end{eqnarray}
which combines a permutation and a reflection.


It is evident from the definition of the symmetric square root of the metric that there is a transformation rule for it, at least between admissible coordinate systems.  That fact is illustrated in the `commutative diagram' (with factors of $diag(-1,1,1,1)$ suppressed):
$$\begin{CD}
g^{\prime} @<tensor << g\\
@VrootVV @VrootVV \\
r^{\prime} @<?<< r
\end{CD}$$

But what can be said, if anything, that is explicit and practical about the transformation from $r_{\mu\nu}$ to $r^{\prime}_{\mu\nu}$, labeled as ``?'' in the diagram? 
  To get a fully explicit transformation rule, at least when the perturbative expansion exists, one can write (schematically)
\begin{eqnarray}
r^{\prime} = \sqrt{g^{\prime} \eta } \eta = \sqrt{ \left( \frac{ \partial x}{ \partial x^{\prime} } \right)^\intercal  r \eta r \frac{ \partial x }{ \partial x^{\prime} } \eta } \eta \end{eqnarray}
and  use $\left(\frac{  \partial x}{ \partial x^{\prime} } \right)^\intercal  r \eta r \frac{ \partial x }{ \partial x^{\prime} } \eta - I$ as the perturbation in the binomial series expansion. 
In detail, the transformation rule is
\begin{eqnarray}
r^{\prime}_{\mu\nu} =  \sqrt{  \frac{ \partial x^{\alpha} }{ \partial x^{\mu\prime} }   r_{\alpha\beta} \eta^{\beta\gamma} r_{\gamma\rho} \frac{ \partial x^{\rho} }{ \partial x^{\sigma\prime} } \eta^{\sigma\delta}  } \eta_{\delta\nu}. \end{eqnarray}
 Thus the transformation rule, at least in the perturbative context, is an infinite series in even powers of $r$.  The series expansion chooses the root near the identity, that is, near $diag(-1,1,1,1);$ presumably one  wants  the `positive' principal square root in all other contexts also. Bilyalov's generalized eigenvector formalism  works more  generally \cite{BilyalovSpinors}, but still requires that  `time' be listed  first among the coordinates; otherwise negative eigenvalues can appear, yielding a complex square root of the metric.  Note that such coordinate restrictions do not arise in the positive definite case considered by Gilbert and Murray \cite{GilbertMurray} and by Bourguignon and Gauduchon \cite{Bourguignon}, which seem to be rather rare examples of the consideration of the symmetric square root of the metric by mathematicians.

Viewed as machines with slots, nonlinear geometric objects have the somewhat peculiar feature of having, one might say, `internal' slots as well as external ones.  The symmetric square root of the metric $r_{\mu\nu}$ has two indices, but its nonlinear transformation rule implies a far more intimate relationship with the basis vectors than one finds for the metric tensor, for which the basis vectors are purely external.  For the metric one can write 
$g_{\mu\nu} = 
 \utilde{\utilde{g}}  (\frac{ \vec{\partial}}{\partial x^{\mu}}, \frac{ \vec{\partial}}{\partial x^{\nu}}) =  
(\frac{ \vec{\partial}}{\partial x^{\mu}}) \cdot  \utilde{\utilde{g}} \cdot \frac{ \vec{\partial}}{\partial x^{\nu}}$: feeding basis vectors into the machine means merely placing them on the outside of the tensor and taking the inner product. Thus one goes from the invariant tensor machine $\utilde{\utilde{g}}$ to its (coordinate) basis components $g_{\mu\nu}$.  It is comparably easy to reverse the procedure using the dual basis $\utilde{d}x^{\mu}$ to get from the components to the invariant tensor machine:
\begin{eqnarray}
\utilde{d}x^{\mu} g_{\mu\nu}  \utilde{d}x^{\nu}  = \utilde{d}x^{\mu} \frac{ \vec{\partial}}{\partial x^{\mu}} \cdot  \utilde{\utilde{g}} \cdot \frac{ \vec{\partial}}{\partial x^{\nu}} \utilde{d}x^{\nu} = \utilde{ \vec{I} } \cdot \utilde{\utilde{g}} \cdot \utilde{ \vec{I} }=
\utilde{\utilde{g}}. 
\end{eqnarray}

One can try to devise an  explicit analogous construction for nonlinear geometric objects like the square root of the metric $r_{\mu\nu}$. 
The result would seem to be the following:  
\begin{eqnarray}
\bold{r} \stackrel{?}{=} c_0 \eta + c_1 \utilde{\utilde{g}}   + c_2 \utilde{\utilde{g}}  \eta \utilde{\utilde{g}} 
+ c_3 \utilde{\utilde{g}}  \eta \utilde{\utilde{g}} \eta \utilde{\utilde{g}}  + \ldots 
\end{eqnarray}
for appropriate coefficients $c_i$ from the perturbation expansion. 
 Here the 0th term has no slots, the first term has two external slots, the second term has two external slots and, in addition, two internal slots that must be summed over, the third term has two external slots and also four internal slots to be summed over, \emph{etc.}, \emph{ad infinitum}.  Viewed as a machine, $\bold{r}$ has two external slots and infinitely many internal slots.  Unfortunately this expression diverges.   
But one can rewrite the binomial series expansion above a bit more abstractly:
\begin{eqnarray*} \bold{r} =
 \sum_{k=0}^{\infty} \frac{ \frac{1}{2}! }{ (\frac{1}{2}-k)! k! } ( \utilde{\utilde{g}} \eta - I)^k \eta. 
\end{eqnarray*}
Once again the invariant machine $\bold{r}$, now defined convergently, has two external slots and infinitely many internal slots. It is straightforward to provide analogous expressions 
 for the square root of the inverse metric $\vec{\vec{g}},$   $\hat{\bold{r}}$ for the square root of the conformal part of the metric, \emph{etc.} 

 From the component standpoint, by contrast, nonlinear geometric objects are technically intricate in comparison to linear ones, but not essentially different or  ineffable.  Are nonlinear geometric objects worth the trouble to define?  While they do sometimes appear in applications \cite{OP,TyutinMass}, at least some of these applications involve only ``quasilinear'' geometric objects---that is, nonlinear geometric objects equivalent to linear ones \cite[p. 79]{AczelGolab}---and so could in principle be carried out using only linear geometric objects.  The role of nonlinear geometric objects in the most economical treatment of spinors is a much better, and indeed compelling,  reason to study them.

 Thus $r_{\mu\nu}$ is (almost) a nonlinear geometric object, where the ``almost'' qualification depends on what coordinate charts one admits. It is, or was, a familiar claim in geometric object theory that one must specify the admissible coordinate systems \cite{KucharzewskiKuczma}.  Given such wiggle room in a vague definition, one could regard $r_{\mu\nu}$ as a geometric object. But above the more common requirement \cite{Nijenhuis,Trautman} that every coordinate system receive a set of components was employed.  By that standard  
$r_{\mu\nu}$ is not a geometric object due to the restriction on admissible coordinates. While unfamiliar, this feature is not a problem.  The custom of admitting all coordinates is an entrenched habit that makes sense when no motivation for a restriction presents itself.  But it is not required for any important task in differential geometry, such as covering a manifold, defining Lie differentiation, or whatever else there might be.  The whole of classical differential geometry with a Lorentzian metric would survive perfectly intact if the restriction to coordinates such that $g_{\mu\nu} \eta^{\nu\alpha}$ have no negative eigenvalues were introduced from the start.

This phenomenon points to the methodological utility of mathematical vagueness, which value is not always appreciated. Some give and take is needed in figuring out what definitions are appropriate in view of the objects to which one wishes to apply them. (This point is somewhat akin to taking a Bayesian view of mathematics \cite{CorfieldMathematics} or crafting a human-faced Bayesianism \cite{HackingRealisticProbability}.)   Imagine a  rigorous mathematician with precise definitions and deductively valid arguments,  who requires geometric objects to admit all coordinates (as usual).  The mathematician then proves, let us suppose, that there is no symmetric square root of the metric that is a geometric object.   Such a result is, strictly speaking, true; it also might seem rather important, much as the Weyl-Cartan theorem seemed quite important.  But the sensible conclusion to draw is that geometric objects as defined  are not quite the right tool for the job.  The problem is not that  $r_{\mu\nu}$ is defective, but that the definition of geometric object was wrong, that is, inappropriate for the intended objects of study. There isn't anything important  (other than admitting arbitrary coordinates, the importance of which is at issue) that one could do with a geometric object that one cannot do with  $r_{\mu\nu}$, which has a Lie derivative in virtue of its transformation properties near the identity. 
 But if one is overly impressed with precision and deductive rigor and employs them in a context where they turn out to be premature, one might charge ahead with the usual definition and fail to notice that the theorem holds due to a minor technical point irrelevant to the substance of classical differential geometry. Schouten, coinventor of the affine connection, held that most mathematicians (except a few ``children of the gods'') make poor physicists, because they pay too much attention to beautiful constructions and too little attention to experiments.  He claimed not to have ``tried to do any physics himself'' regarding the problem at hand, involving conformal transformations, ``[b]ecause he is convinced that this is a job only physicists can do properly.'' \cite{SchoutenMesonConformal} While his point is not exactly mine, clearly Schouten would have recognized that it would be a mistake for physicists and philosophers to think that progress in mathematics was automatically progress in physics and philosophy, because mathematicians have  different aims and criteria. 
Classical component-based differential geometry has some useful vagueness here.  Imposing a restriction on admissible coordinates in a classical context is a much smaller change, and more obviously irrelevant to every result in classical differential geometry,  than is the corresponding modern task.

Of course there \emph{is} a role for mathematical rigor; sometimes one really needs to know which results are proven theorems.  There are also occasions where one needs to treat problems globally, at least in General Relativity, on which occasions introducing local coordinates is clearly unhelpful.  But the principle of being able to walk before running applies: one needs to have the local problem under control before tackling the global problem.  Surprisingly enough, the local spinor problem was not understood properly (without a lot of surplus structure and a gratuitous gauge group) by anyone (except maybe the precursors noted above) prior to 1965, notwithstanding authoritative statements from Weyl, Cartan, \emph{inter alia}, and remains so even for most authors today.  Thus the application of precision and deductive rigor turns out to have been premature.  



\subsection{Are Coordinates Surplus Structure?}

The reader might have noticed a contrast between some attitudes suggested here and the common view that coordinates are an annoying piece of surplus structure now already successfully overcome and eliminated by modern differential geometry. 
This modern view is easier to hold if one follows the common practice of failing to study nonlinear geometric objects.  
 But coordinates are no panacea; the rise of modern techniques was not a fall from classical paradise.  Coordinates are an occasionally tedious but quite flexible bookkeeping device. 
Of course there is no point in introducing coordinate basis components in contexts that are already handled as well or better without them.  
 Global results for topologically nontrivial manifolds will be difficult or impossible to come by without modern `coordinate-free' techniques.  Nontrivial bundles over topologically trivial manifolds, which might arise even in alternative gravitational theories where space-time is globally quite boring,  also benefit.

 Whether one ought to avoid coordinates and when depends on whether one can do without them everything important that one can do with them,  whether doing without them requires introducing an alternative irritant that might be worse, and whether coordinates can be \emph{introduced} or are \emph{already} present and hence unavoidable at some level.  In the present state of mathematics, at least, the following statements appear to be true.  First, coordinates are \emph{already} present in the usual definition of a manifold and thus apparently are unavoidable, hence not yet shown to be surplus structure.
(It seems unhelpful to avoid talking about coordinates by refusing to assign meaning to a word---``manifold,'' ``smooth,'' or the like---that traditionally has a precise and important meaning.) Second, refusing to talk about coordinates either blocks the introduction of a Lie derivative or requires active diffeomorphisms. Third, active diffeomorphisms presuppose a substantival notion of manifold point individuation at odds with that required to free General Relativity from indeterminism along the lines of Einstein's relational-flavored point-coincidence argument as a response to the hole argument (see, \emph{e.g.}, \cite{Norton}). 
 If coordinates cannot be avoided without giving up manifolds, if Lie differentiation requires mention of either coordinates or active diffeomorphisms, and if active diffeomorphisms require an un-General Relativistic notion of point identity that motivates indeterminism, then it is by no means yet clear that one always ought to try to avoid coordinates.

Defining Lie derivatives requires that something move---either coordinates move relative to space-time points (passive coordinate transformations), or fields move relative to space-time points (active diffeomorphisms). Something needs to change in order to \emph{represent} the group(oid) of (more or less) arbitrary transformations.  If nothing changes, then \emph{a fortiori} nothing changes like a vector, or a 3-form, or the like. In a few simple cases one might hit upon the Lie derivative even without any unifying principle---\emph{e.g.}, the commutator of two vectors; but neither the full meaning of the construction nor anything like adequate generality would be realized.   One sometimes sees treatments of differential geometry that deal with invariant bold-faced objects (largely avoiding coordinates), modeled after vector calculus,  and also  avoid active  mappings \cite{YoungVectorTensor,Simmonds}.   In such work covariant differentiation appears to arise due to ordinary partial differentiation of an invariant entity built using both components and coordinate basis vectors;  differentiation of the coordinate basis vectors is supposed to give rise to the Christoffel symbol terms. That seems very satisfactory.  But that cannot be what is really happening, because the connection is an independent field; evidently  one already knows in advance what the basis vectors are, namely, coordinate gradients of the position vector ${\bf \vec{x} },$ which involves reference to a flat metric.   Of course ${\bf \vec{x} }$ does not exist relative to a curved metric, so the technique is rather restricted in application.  
If neither the coordinates nor the fields move, there evidently is no concept of a Lie derivative, and hence no notion of a Killing vector field or the like, and no adequate way to treat space-time transformations (on curved manifolds), which are associated with conserved quantities \cite{BergmannConservation}.  That being unacceptable, one needs for something to move, either coordinates in a passive transformation of labels, or fields in an active diffeomorphism.

If one treats OP spinors using active diffeomorphisms, one will encounter the analog of the coordinate restriction to keep the components of $g_{\mu\nu} \eta^{\nu\alpha}$ from having negative eigenvalues.  Thus \emph{arbitrary} diffeomorphisms, though customary, are excluded.  One could learn to live with that change.  But parts of  modern mathematical literature would need to be reworked to avoid error or irrelevance.  By contrast, in classical coordinate techniques, it is obvious that complete generality of coordinates, as opposed to arbitrariness near the identity, for example, does little or no  work.

Active diffeomorphisms presuppose a mathematical metaphysics of point individuation of a rather absolutist or substantivalist sort. Mathematical points have haecceities, primitive identities not tied to what happens at the points.  One can mail the fields elsewhere while leaving the points behind; the points don't go with the fields.  That is almost precisely the \emph{denial} of the point of Einstein's point-coincidence argument, which played a crucial role in rescuing generally covariant field equations from indeterminism.  
Thus one is left to worry about indeterminism again, as in the 1980s resurfacing of Einstein's hole argument \cite{HoleStory}.  While various strategies for dealing with the hole argument have been deployed, a natural and largely satisfactory approach is simply to refuse to build mathematics on a metaphysical foundation at odds with that of our  best current theory of gravity, General Relativity.  General Relativity being the spur that made differential geometry so dominant a part of 20th century mathematics, this tacit repudiation of its lessons regarding point individuation is ironic, driving a wedge between mathematical points and physical space-time points.  

Mathematicians as such need not worry much about the plausibility of their manifold point individuation assumptions. There are plenty of mathematically interesting manifolds that are not even intended, by virtue of dimension, metric signature, lack of a metric, or something else, to pertain to space-time. Perhaps only if mathematicians wish for their manifolds to pertain to space-time does the metaphysics of manifold point individuation rise above the level of convention.  At that point they become in effect physicists and/or philosophers and new criteria emerge.  If one can find a treatment that avoids the surplus structure of a misleading metaphysics of points and  that allows one to express and calculate everything that one needs without using coordinates, then the promise of modern differential geometry for physicists and philosophers will be achieved.

%

\subsection{Conformal  Group Yields Linear Transformation Law}

Besides the full nonlinear transformation law for (nearly) general coordinate transformations, it is of interest to ascertain under what circumstances the transformation law is linear---that is, to find out what is the stabilizer group.  The answer turns out to be the 15-parameter ($\frac{(n+1)(n+2)}{2}$ in $n$ dimensions)  conformal group, as will now appear.  That group contains the Poincar\'{e} group of special relativistic field theory, a fact that explains why one only sees a linear transformation law in special relativistic field theory in Cartesian coordinates.  Besides the very familiar four ($n$) rigid space-time translations and six ($\frac{n(n-1)}{2}$) familiar Lorentz transformations (three ($\frac{n^2 -3n +2}{2}$) rotations and three ($n-1$) boosts), the conformal group \cite{FultonRohrlichWitten} involves a scale transformation $ x^{\mu\prime} = s x^{\mu}$ (where $s$ is a real number, nonzero and positive without loss of generality) and four ($n$) ``acceleration'' transformations
\begin{eqnarray}
x^{\mu\prime}  = \frac{ x^{\mu} + a^{\mu}  }{1 + 2 a^{\mu} \eta_{\mu\nu} x^{\nu} + a^2 x^2}, 
\end{eqnarray}
where indices are moved with the matrix $diag(-1,1,1,1)$  and $a^2 = a^{\mu} \eta_{\mu\nu} a^{\nu},$ $x^2=x^{\mu} \eta_{\mu\nu} x^{\nu}.$ (In $n$ dimensions, one can append or remove $1$'s and/or $-1$'s along the diagonal as needed.)  
  These groups are formal in the sense that, whether or not the initial coordinates are approximately Cartesian, the same functional form of the transformation is used.   
Making contact with particle physicists' work on nonlinear group representations, which grew partly out of the OP spinor formalism  (see also \cite{IshamSalamStrathdee,BorisovOgievetskii,ChoFreund,Miglietta,NeemanSijackiAnnals}), one can call the conformal group the stability group for the nonlinear representation.  The symmetric square root of the metric is thus a nonlinear representation of  the general coordinate transformation group (dividing out the  coordinate reorderings involving time), with  the conformal group as its stabilizer.  This  example of a nonlinear geometric object (apart from coordinate restrictions) is equivalent to a linear geometric object, namely, the metric (or, for that matter, its inverse, or various densitized metrics) and so is  quasilinear. 
  Quasilinear geometric objects have the advantage of providing a bridge between linear cases where everything is familiar, such as how to take a Lie or (when defined) covariant derivative, and nonlinear cases where much is different.

It is plausible by inspection, and rigorously true, that the conformal group is exactly the group that leaves invariant the conformal metric density $\hat{\eta}_{\mu\nu}$ (which determines the null cones), if one pretends to have a conformally flat null cone structure in the theory.  (There is no such structure here, only the matrix $diag(-1,1,\ldots,1),$ but a useful trick results nonetheless.)   This fact, along with strategic insertion of identities in the nonlinear transformation law, will yield the linear transformation law for the conformal group.  From above, the nonlinear transformation rule is
\begin{eqnarray}
r^{\prime}_{\mu\nu} =  \sqrt{  \frac{ \partial x^{\alpha} }{ \partial x^{\mu^{\prime}} }   r_{\alpha\beta} \eta^{\beta\gamma} r_{\gamma\rho} \frac{ \partial x^{\rho} }{ \partial x^{\sigma^{\prime}} } \eta^{\sigma\delta}  } \eta_{\delta\nu} = \nonumber \\
 \sqrt{  \frac{ \partial x^{\alpha} }{ \partial x^{\mu^{\prime}} }   r_{\alpha\beta}  \frac{\partial x^{\beta} }{\partial x^{\lambda^{\prime}} }     \frac{\partial x^{\lambda^{\prime}} }{\partial x^{\chi } }       \eta^{\chi\gamma} r_{\gamma\rho} \frac{ \partial x^{\rho} }{ \partial x^{\sigma^{\prime}} } \eta^{\sigma\delta}  } \eta_{\delta\nu} = \nonumber \\
\sqrt{  \frac{ \partial x^{\alpha} }{ \partial x^{\mu^{\prime}} }   r_{\alpha\beta}  \frac{\partial x^{\beta} }{\partial x^{\lambda^{\prime}} }     \frac{\partial x^{\lambda^{\prime}} }{\partial x^{\chi } }       \eta^{\chi\gamma}    \frac{\partial x^{\epsilon^{\prime}} }{\partial x^{\gamma} }     \frac{\partial x^{\phi} }{\partial x^{\epsilon^{\prime} } }      r_{\phi\rho} \frac{ \partial x^{\rho} }{ \partial x^{\sigma^{\prime}} } \eta^{\sigma\delta}  } \eta_{\delta\nu} = \nonumber \\
\sqrt{  \frac{ \partial x^{\alpha} }{ \partial x^{\mu^{\prime}} }   r_{\alpha\beta}  \frac{\partial x^{\beta} }{\partial x^{\lambda^{\prime}} }   \left|  \frac{\partial x}{\partial x^{\prime} }   \right|^{-\frac{2}{n}}   \left|  \frac{\partial x}{\partial x^{\prime} }   \right|^{\frac{2}{n}}   \frac{\partial x^{\lambda^{\prime}} }{\partial x^{\chi } }       \eta^{\chi\gamma}    \frac{\partial x^{\epsilon^{\prime}} }{\partial x^{\gamma} }     \frac{\partial x^{\phi} }{\partial x^{\epsilon^{\prime} } }      r_{\phi\rho} \frac{ \partial x^{\rho} }{ \partial x^{\sigma^{\prime}} } \eta^{\sigma\delta}  } \eta_{\delta\nu} = \nonumber \\
\sqrt{  \frac{ \partial x^{\alpha} }{ \partial x^{\mu^{\prime}} }   r_{\alpha\beta}  \frac{\partial x^{\beta} }{\partial x^{\lambda^{\prime}} }   \left|  \frac{\partial x}{\partial x^{\prime} }   \right|^{-\frac{1}{n}} \left(  \left|  \frac{\partial x}{\partial x^{\prime} }   \right|^{\frac{2}{n}}   \frac{\partial x^{\lambda^{\prime}} }{\partial x^{\chi } }       \eta^{\chi\gamma}    \frac{\partial x^{\epsilon^{\prime}} }{\partial x^{\gamma} }  \right)   \frac{\partial x^{\phi} }{\partial x^{\epsilon^{\prime} } }      r_{\phi\rho} \frac{ \partial x^{\rho} }{ \partial x^{\sigma^{\prime}} }   \left|  \frac{\partial x}{\partial x^{\prime} }   \right|^{-\frac{1}{n}}  \eta^{\sigma\delta}  } \eta_{\delta\nu}. 
\end{eqnarray}
The quantity in parentheses, as a set of numbers, is formally equal to the components of  a flat inverse conformal metric density $\hat{\eta}^{\mu\nu},$ a tensor density of weight $\frac{1}{2}$ (or more generally, $\frac{2}{n}$), under a transformation from Cartesian coordinates to some new coordinates. (This is just a formal  mathematical trick, because only the numerical matrix $diag(-1,1,1,1)$ is present!) But in the special case where the transformation is in the (15- or $\frac{(n+1)(n+2)}{2}$-parameter)  group of conformal transformations, the parenthetic quantity simplifies dramatically:
\begin{eqnarray}
  \left|  \frac{\partial x}{\partial x^{\prime} }   \right|^{\frac{2}{n}}   \frac{\partial x^{\lambda^{\prime}} }{\partial x^{\chi } }       \eta^{\chi\gamma}    \frac{\partial x^{\epsilon^{\prime}} }{\partial x^{\gamma} }   = \eta^{\lambda\epsilon}  \end{eqnarray}
(with no primes because conformal transformations are symmetries of the entity in question), where   $\eta^{\lambda\epsilon} $
is just the matrix $diag(-1,1,\ldots,1)$ as usual.  Installing this simplification into the transformation rule, one encounters the square root of a perfect square:
\begin{eqnarray}
r^{\prime}_{\mu\nu} =  \nonumber \\  \sqrt{  \frac{ \partial x^{\alpha} }{ \partial x^{\mu^{\prime}} }   r_{\alpha\beta}  \frac{\partial x^{\beta} }{\partial x^{\lambda^{\prime}} }   \left|  \frac{\partial x}{\partial x^{\prime} }   \right|^{-\frac{1}{n}} \left(  \left|  \frac{\partial x}{\partial x^{\prime} }   \right|^{\frac{2}{n}}   \frac{\partial x^{\lambda^{\prime}} }{\partial x^{\chi } }       \eta^{\chi\gamma}    \frac{\partial x^{\epsilon^{\prime}} }{\partial x^{\gamma} }  \right)   \frac{\partial x^{\phi} }{\partial x^{\epsilon^{\prime} } }      r_{\phi\rho} \frac{ \partial x^{\rho} }{ \partial x^{\sigma^{\prime}} }   \left|  \frac{\partial x}{\partial x^{\prime} }   \right|^{-\frac{1}{n}}  \eta^{\sigma\delta}  } \eta_{\delta\nu} = \nonumber \\
\sqrt{  \frac{ \partial x^{\alpha} }{ \partial x^{\mu^{\prime}} }   r_{\alpha\beta}  \frac{\partial x^{\beta} }{\partial x^{\lambda^{\prime}} }     \left|  \frac{\partial x}{\partial x^{\prime} }   \right|^{-\frac{1}{n}}   (\eta^{\lambda\epsilon})  \frac{\partial x^{\phi} }{\partial x^{\epsilon^{\prime} } }      r_{\phi\rho} \frac{ \partial x^{\rho} }{ \partial x^{\sigma^{\prime}} }    \left|  \frac{\partial x}{\partial x^{\prime} }   \right|^{-\frac{1}{n}}     \eta^{\sigma\delta}  } \eta_{\delta\nu} = \nonumber \\
  \frac{ \partial x^{\alpha} }{ \partial x^{\mu^{\prime}} }   r_{\alpha\beta}  \frac{\partial x^{\beta} }{\partial x^{\lambda^{\prime}} }   \left|  \frac{\partial x}{\partial x^{\prime} }   \right|^{-\frac{1}{n}}   \eta^{\lambda\delta} \eta_{\delta\nu}  = \nonumber \\
\frac{ \partial x^{\alpha} }{ \partial x^{\mu\prime} }   r_{\alpha\beta}  \frac{\partial x^{\beta} }{\partial x^{\nu\prime} }    \left|  \frac{\partial x}{\partial x^{\prime} }   \right|^{-\frac{1}{n}}:
\end{eqnarray}
the symmetric square root of the metric transforms as a tensor density of a certain weight under the conformal group, and hence as a tensor under the Poincar\'{e} group.  (Note that for spherical polar coordinates, for example, the perturbative expansion fails, but the transformation nonetheless makes sense, as one sees easily by inspection.)

The transformation rule for the symmetric square root of the conformal metric density $\hat{r}_{\mu\nu} = r_{\mu\nu} \sqrt{-g}^{\; -\frac{1}{n} }$ is similar, but with modest changes to ensure preservation of the anti-unit determinant condition $|\hat{r}_{\mu\nu}|=-1.$  For general coordinate transformations one has
\begin{eqnarray}
\hat{r}^{\prime}_{\mu\nu} =  \sqrt{  \left|  \frac{\partial x}{\partial x^{\prime} }   \right|^{-\frac{2}{n}}  \frac{ \partial x^{\alpha} }{ \partial x^{\mu^{\prime}} }   \hat{r}_{\alpha\beta} \eta^{\beta\gamma} \hat{r}_{\gamma\rho} \frac{ \partial x^{\rho} }{ \partial x^{\sigma^{\prime}} } \eta^{\sigma\delta}  } \eta_{\delta\nu}.
\end{eqnarray}
For the conformal group one has the simplification 
\begin{eqnarray}
\hat{r}^{\prime}_{\mu\nu} =    \nonumber \\ \sqrt{ \left|  \frac{\partial x}{\partial x^{\prime} }   \right|^{-\frac{2}{n}} \frac{ \partial x^{\alpha} }{ \partial x^{\mu^{\prime}} }   \hat{r}_{\alpha\beta}  \frac{\partial x^{\beta} }{\partial x^{\lambda^{\prime}} }   \left|  \frac{\partial x}{\partial x^{\prime} }   \right|^{-\frac{2}{n}} \left(  \left|  \frac{\partial x}{\partial x^{\prime} }   \right|^{\frac{2}{n}}   \frac{\partial x^{\lambda^{\prime}} }{\partial x^{\chi } }       \eta^{\chi\gamma}    \frac{\partial x^{\epsilon^{\prime}} }{\partial x^{\gamma} }  \right)   \frac{\partial x^{\phi} }{\partial x^{\epsilon^{\prime} } }      \hat{r}_{\phi\rho} \frac{ \partial x^{\rho} }{ \partial x^{\sigma^{\prime}} } \eta^{\sigma\delta}  } \eta_{\delta\nu}  \nonumber \\ =
\sqrt{  \frac{ \partial x^{\alpha} }{ \partial x^{\mu^{\prime}} }   \hat{r}_{\alpha\beta}  \frac{\partial x^{\beta} }{\partial x^{\lambda^{\prime}} }    \eta^{\lambda\epsilon}   \left|  \frac{\partial x}{\partial x^{\prime} }   \right|^{-\frac{4}{n} }   \frac{\partial x^{\phi} }{\partial x^{\epsilon^{\prime} } }      \hat{r}_{\phi\rho} \frac{ \partial x^{\rho} }{ \partial x^{\sigma^{\prime}} } \eta^{\sigma\delta}  } \eta_{\delta\nu}  = \nonumber \\
  \frac{ \partial x^{\alpha} }{ \partial x^{\mu^{\prime}} }   \hat{r}_{\alpha\beta}  \frac{\partial x^{\beta} }{\partial x^{\nu^{\prime}} }      \left|  \frac{\partial x}{\partial x^{\prime} }   \right|^{-\frac{2}{n}}_,
\end{eqnarray}
a tensor density of weight $-\frac{2}{n}$ under the conformal group; this is just the weight required to preserve the anti-unit determinant condition  $|\hat{r}_{\mu\nu}|=-1.$  One notices that  
$\hat{r}_{\mu\nu},$ or rather its inverse $\hat{r}^{\mu\nu},$  is the quantity to which spinors most readily couple, once one has pared away the fat of the asymmetric part of the tetrad and the volume element.  They are usually retained as though they couldn't be eliminated, giving 16 components ($n^2$) where only 9 components ($\frac{n^2 + n -2}{2}$) are needed, an excess of 7 pieces of surplus structure in 4 dimensions or $\frac{n^2 -n + 2}{2} $ extra pieces in general.  Evidently modern aspirations to avoid surplus structure in differential geometry have not been fulfilled in the more common spinor formalism. Ogievetsky and Polubarinov noted only the linearity under the Poincar\'{e} group \cite{OPspinorReprint}, not the full 15-parameter conformal group, but in Ogievetsky's later work the conformal group increasingly appeared \cite{OgievetskyLNC,BorisovOgievetskii}.


\subsection{Linear Spinor   Transformation  for Conformal Group}

Above it has been shown that the symmetric square root of the (or its inverse) metric has a transformation law that, for the special case of the conformal group, is linear.  The square root of the metric is interesting, for our purposes, not primarily as an end in itself; after all, it is equivalent as a geometric object (apart from the question of admissible coordinate systems) to the metric, that is, quasilinear in the sense of Aczel and Golab \cite{AczelGolab}.  (It also appears in various nonlinear gauge theories \cite{LPTT,Kirsch,AliCapozzielloNonlinearGroup}.)  
Rather, the square root of the metric is useful for coupling spinors to a curved metric; at that point one leaves the realm of quasilinearity for an essentially nonlinear geometric object, or at any rate an entity with an essentially nonlinear transformation law (along with coordinate restrictions and spinorial two-valuedness).  

That the spinorial coordinate transformation rule is linear for the 15-parameter conformal group can be seen in two ways.  One way, suitable for infinitesimal transformations, is to use the explicit form of the Lie derivative given by Ogievetsky and Polubarinov \cite{OPspinorReprint}.  This expression involves, as one would expect, a transport term involving the generating vector field and the coordinate gradient of the spinor, a term that specifies a Lorentz transformation out of the curl (the antisymmetric part of the gradient) of the generating vector field, and a novel  correction term involving the   symmetrized gradient of the generating vector field, which term depends nonlinearly on the symmetric square root of the metric.  While initially forbidding, this novel term, on continued contemplation, gives up some of its secrets.  The new term, which is what makes it possible to take the Lie derivative of a spinor field (which, one recalls, is widely believed to be impossible),  involves a commutator of the square root of the metric with the symmetrized coordinate gradient of the generating vector field.  Lie differentiation for spinors with respect to an arbitrary vector field is not impossible, nor even fundamentally different from Lie differentiation for (nonlinear) geometric objects \cite{Yano}; it is merely difficult and unfamiliar.  On account of the commutator, the new term vanishes if the symmetrized gradient is  proportional to the identity matrix, or, giving up their use of an imaginary time coordinate, proportional to $\eta^{MN}.$  But that condition is simply the conformal Killing equation, if one treats the matrix $\eta^{MN}$ formally as an inverse metric.  Hence the conformal Killing vector fields of the `metric' $\eta^{MN}$ (not $\hat{g}^{\mu\nu}$) annihilate the correction term.

If one wishes to give the spinor density weight $w$ ($\frac{3}{8}$ being the proper weight to achieve a conformally invariant Dirac operator in four space-time dimensions), then an extra term  $+ w \psi \xi^{\mu},_{\mu}$ appears, as is usual for Lie derivatives of densities \cite{Schouten,OPspinorReprint,Anderson,Israel}; linearity is not threatened.   

For \emph{finite} conformal transformations (even those not connected to the identity), in the sense of coordinate transformations such that there exists some (positive) function $F$ such that
\begin{eqnarray} 
\frac{ \partial x^{\mu^{\prime} } }{ \partial x^{\alpha} } \eta^{AB} \frac{ \partial x^{\nu^{\prime} } }{ \partial x^{\beta} }    = \eta^{MN} F,
\end{eqnarray}
the transformation law is still linear.  One can factor out the determinant to get
\begin{eqnarray} 
\frac{ \partial x^{\mu^{\prime} } }{ \partial x^{\alpha} }  \left| \frac{\partial x^{\prime} }{\partial x}\right|^{-\frac{1}{n}}  \eta^{AB}   \left| \frac{\partial x^{\prime} }{\partial x}\right|^{-\frac{1}{n}}   \frac{ \partial x^{\nu^{\prime} } }{ \partial x^{\beta} }    = \eta^{MN}:
\end{eqnarray}
$\frac{ \partial x^{\mu^{\prime} } }{ \partial x^{\alpha} }  \left| \frac{\partial x^{\prime} }{\partial x}\right|^{-\frac{1}{n}} $ is therefore a Lorentz transformation (possibly varying from point to point).  The whole conformal transformation is therefore the product of a local  stretching and a local Lorentz transformation.  Making such a transformation on a tetrad thus stretches and Lorentz-transforms the world (coordinate, Greek) index in accord with the vector (or covector) coordinate transformation law.  Assuming that the tetrad was initially symmetric (so as to give a transformation of the square root of the metric under coordinate transformations), one can recover symmetry by making \emph{the same} local Lorentz transformation on the local Lorentz (Latin) index, so that both indices receive the same treatment.  It does not matter what the value of the square root of the metric is; only its being symmetric matters.  The local Lorentz transformation acting on the Latin index is thus determined wholly by the coordinate transformation, \emph{not} by the  (conformal part of the) metric.  Thus nonlinearity is averted for the transformation of the symmetric square root of the metric for these conformal coordinate transformations.

Whereas the spinors accompanying a tetrad are coordinate scalars and are spinors under (the double cover of) a distinct local $O(1,3)$ group, Ogievetsky-Polubarinov spinors have a distinctive spinorial coordinate transformation law that reduces to the usual one in particle physics in appropriate special cases.  One can derive the Ogievetsky-Polubarinov spinor transformation law by fixing the local $O(1,3)$ gauge freedom (given a coordinate system) to make the tetrad components symmetric (after an index is moved with $\eta_{MN}$); the corresponding spinor transformation for that element of the local $O(1,3)$ group is then applied to the spinor, giving a coordinate- and metric-dependent boost-rotation to the spinor as its coordinate transformation law.   For the special case of the conformal group of coordinate transformations, because the need to preserve the symmetry of the tetrad (that is, the square root of the metric) determines the local Lorentz transformation that determines (up to a sign) what happens to the spinor field, it follows that the spinor transformation is also independent of the metric, and is merely linear in the spinor.  That is just what one expects from experience with spinors in Cartesian coordinates in Special Relativity, where one makes only Poincar\'{e} transformations \cite{Kaku},  a subgroup of this conformal group.

\subsection{Necessity and Sufficiency of Coordinate Reordering:  Explicit Treatment in Two Dimensions}

One powerful way to take the symmetric square root of the metric is to use a generalized eigenvalue formalism, where one has both the metric $g_{\mu\nu}$ and a Lorentzian matrix $\eta_{\mu\nu}=diag(-1,1,1,1)$  \cite{BilyalovConservation,NullCones1}. 
It is convenient, though not necessary, to factor out the volume element $g=_{def} det(g_{\mu\nu})$ by working with the conformal metric density $\hat{g}_{\mu\nu}$, which has determinant $-1$.  
Due to the indefiniteness of both  $g_{\mu\nu}$ and $\eta_{\mu\nu},$ the usual theorems about a complete set of real eigenvalues with orthogonal and complete eigenvalues do not apply in general.  

There are several things that can go `wrong' in this generalized eigenvalue formalism, compared to the familiar case where one of the matrices is positive definite or one takes the ordinary eigenvalues of a symmetric matrix. One is that the eigenvalues need not be real.  Another is that there need not be a complete set of eigenvectors due to nontrivial Jordan blocks.  In four dimensions, one can have four real eigenvalues with four independent eigenvectors, or four real eigenvectors with three independent eigenvectors, or four real eigenvalues with two independent eigenvectors, or two real and two complex conjugate eigenvalues with two real and two complex conjugate eigenvectors  \cite{PetrovEinsteinSpaces,HallArab,HallDiff,HallNegm,Hall5d,BilyalovConservation,KopczynskiTrautman,BilyalovSpinors,NullCones1}. 
It turns out that neither complex eigenvalues nor nontrivial Jordan blocks with the associated shortage of eigenvectors cause any problem.  The only problem arises if some eigenvalue(s) is \emph{negative}, because then the square root is not real. (Given the metric signatures, there will be either $0$ or $2$ negative eigenvalues.)  
 Negative real parts for complex eigenvalues do not pose a problem; one can take their square roots to lie in the right half plane and have a positive real sum.  
To find the symmetric square root of the metric, one next finds the eigenvectors, which come as a pair of complex conjugates if the eigenvalues do (along with the real pair from the two suppressed dimensions).  One can find the square root of the metric by taking the square root of the eigenvalues and choosing the  roots to lie in the right half-plane so that the square roots are themselves complex conjugates, and thus obtain  a real symmetric square root  \cite{HighamRoot87,BilyalovConservation,HighamRoot,BilyalovSpinors,HighamSquareRoot}.  As Higham succinctly puts the matter, ``Any matrix with no nonpositive real eigenvalues has a unique square root for which every eigenvalue lies in the open right half-plane.'' \cite{HighamRoot} The matrix for which one wants a square root is  $g_{\mu\nu}\eta^{\nu\alpha}$ or the like, in \emph{each} coordinate system. It is also connected to the identity $\eta_{MN}$, that is, the `positive' or principal square root. Having no negative eigenvalues is not merely sufficient, but also necessary for a real square root determined by the metric:
 ``Let $A \in \mathbb{R}^{n \times n}$ be nonsingular.  If $A$ has a real negative eigenvalue, then $A$ has no real square roots which are functions of A.'' \cite{HighamRoot87}   A square root that is not a function of the original matrix (a tetrad, for example is such a square root of the metric) contains surplus structure. 
  It appears not to be possible to let the square root of the metric go complex (in fact, imaginary) while maintaining useful degrees of differentiability.

%
%

In two dimensions one readily solves for the eigenvalues using the quadratic formula.  Consider the eigenvalue problem 
\begin{eqnarray}
| \hat{g}^{\mu\nu} - \lambda \eta^{\mu\nu} | = 
\left| \left[ \begin{array}{cc}
\hat{g}^{00} -\lambda \eta^{00}  & \hat{g}^{01}  \\ 
\hat{g}^{01}  & \hat{g}^{11} -\lambda \eta^{11}  
\end{array}
\right] \right|  = -\lambda^2 + \eta_{\mu\nu} \hat{g}^{\mu\nu} \lambda -1=0.
\end{eqnarray}
Here $\eta^{\mu\nu}=diag(-1,1).$ 
One readily finds that 
\begin{eqnarray} 
\lambda = \frac{\eta_{\mu\nu} \hat{g}^{\mu\nu}   \pm \sqrt{  (\eta_{\mu\nu} \hat{g}^{\mu\nu})^2  - 4} }{2}.
\end{eqnarray}
One readily sees that if $|\eta_{\mu\nu} \hat{g}^{\mu\nu}|<2,$ then the eigenvalues and hence the eigenvectors are complex, giving a real square root of the (conformal part of the) metric. If $\eta_{\mu\nu} \hat{g}^{\mu\nu} \geq 2,$ then the eigenvalues are positive and again the square root is real.  But if 
$\eta_{\mu\nu} \hat{g}^{\mu\nu} \leq -2,$ then the eigenvalues are negative and the square root is not real, which is bad.  One therefore has, in two space-time dimensions, the following necessary and sufficient condition for admissibility of coordinates:
\begin{eqnarray}
\eta_{\mu\nu} \hat{g}^{\mu\nu} > -2.
\end{eqnarray}
(This condition is equivalent to $\eta^{\mu\nu} \hat{g}_{\mu\nu} > -2,$ as one would hope.)  
This of course holds neighborhood by neighborhood; a coordinate system that violates this condition in only some places is admissible in those places where the condition hold.  
Thus it is not the case that all coordinate systems are admissible, and the general linear group $GL(2, \mathbb{R})$ is not appropriate: it is not the case for admissible coordinates that the derivative of one coordinate system with respect to the other at a point is arbitrary.  Because $\eta_{\mu\nu}$ is indefinite, nonlinear group representations have to care about the minus sign, or, one might say casually, about which coordinate is time.  Thus electrons know what a time coordinate is, as do all other spinorial entities.  The standards for being time are extremely lax, but they are not trivial.  

But one notices that by interchanging the coordinates $x^{0}$ and $x^{1}$, and thus interchanging $g^{00}$ and $g^{11}$, flips the sign of  $\eta_{\mu\nu} \hat{g}^{\mu\nu} $ and hence makes an inadmissible coordinate system admissible.  Thus not any (ordered) coordinate system is admissible, but any \emph{appropriately ordered} coordinate system is admissible.  Such coordinate swapping, one recalls, is part of the service performed by Bilyalov's $T$ matrix \cite{BilyalovConservation} and its relatives. 
In higher dimensions this sort of simple and direct argument has not been given and is not trivial. 
 Bilyalov's proof \cite{BilyalovConservation} takes a somewhat different form.

 Obtaining negative eigenvalues, then, is simply a sign that one has proposed  as a time coordinate what is really (from the perspective of nonlinear geometric objects linear for the conformal subgroup) a spatial coordinate. Merely reordering the coordinates at the point solves the problem.  If the coordinate transformation is affine, so that its gradient is constant, then this move at a point in fact suffices globally.  If the coordinate transformation is not affine, then reordering the coordinates at a point will tend in some cases to reduce the size of admissible coordinate charts, because a coordinate chart that has time in the first position in some parts of space-time and not in the first position in other parts of space-time will need to have the latter regions truncated.  This is really not a problem, though it violates the modern habit of admitting all possible coordinate systems \emph{a priori} and never reconsidering them later, a habit that was not so entrenched in the classical literature on geometric objects \cite{KucharzewskiKuczma}.

One might wonder why this coordinate restriction apparently has gone unnoticed, at least prior to Bilyalov's work.  The answer, or much of it, is that previous writers on nonlinear representations of (a sufficiently large piece of) $GL(4, \mathbb{R})$ that are linear for the Lorentz subgroup have been \emph{content to work near the identity}  \cite{OPspinorReprint,IshamSalamStrathdee,IshamSalamStrathdee2,BorisovOgievetskii,ChoFreund,BorisovInduced1}.  
This has been evident in most cases from their explicitly saying so, their working infinitesimally, or their achieving finite transformations by exponentiating an infinitesimal transformation.


\subsection{Differentiation of Nonlinear Geometric Objects}

Nonlinear geometric objects have not been investigated in much detail by differential geometers or general relativists.  Ironically, particle physicists hit upon the idea of nonlinear group representations at about the same time (1960s) that the newer so-called coordinate-free notion of differential geometry helped to divert mathematicians' interests away from the idea of geometric objects.  Some important results were obtained for Lie and covariant differentiation of nonlinear geometric objects, however  \cite{Tashiro1,Tashiro2,Yano,SzybiakCovariant,SzybiakLie}.
Though typically one figures out the form of the covariant derivative by hand, there is in fact a general formula for figuring out the covariant derivative, when it is defined---which it isn't if the geometric object in question has a transformation rule involving second derivatives, as connections do.  
The same holds for the Lie derivative (which is always defined), including that for a nonlinear geometric object.  Coordinate transformations near the identity suffice \cite{SzybiakCovariant,SzybiakLie}, which explains why the usual (``polar'' \cite{Jackson}) vectors and axial vectors  have the same Lie and covariant derivative formulas, despite doing different things under time reversal or spatial inflection; by the same token, the difference between scalars and pseudoscalars disappears infinitesimally.  More generally, objects which impose some restrictions on coordinate transformations far from the identity nonetheless have well defined Lie and (where appropriate) covariant derivatives.  

For nonlinear geometric objects, a general result is that the Lie and covariant derivatives themselves are not geometric objects, but the Lie (or covariant) derivative forms a geometric object \emph{with} the object itself  \cite{Tashiro1,Tashiro2,Yano,SzybiakCovariant,SzybiakLie}.
 If $\chi$ is a nonlinear geometric object, then 
only the  \emph{pair}
 $\langle \chi, \mathcal{L}_{\xi} \chi \rangle$ is a geometric object---that is, has a transformation rule giving the components in one coordinate system in terms of the components in the other coordinate system--- not $ \mathcal{L}_{\xi} \chi$ alone.  The reason is that the Lie derivative in one coordinate system depends on both the Lie derivative and the original field in another coordinate system.  The same holds for covariant derivatives (unless special circumstances make noxious terms vanish, as will occur in part for OP spinors):
 $\langle \chi, \nabla  \chi \rangle$ is a geometric object, but $\nabla  \chi$ alone is not \cite{Yano,SzybiakCovariant,SzybiakLie}.  An  exceptional case is when $\nabla  \chi$ vanishes, for example. A relevant partial instance is  the spinorial case when one can write $\chi$ as $\langle g_{\mu\nu}, \psi \rangle,$  $\nabla g$ vanishes, and the transformation rule for $\psi$ is linear in $\psi$ (though dependent on $g$), as will appear below.  The fact that an Ogievetsky-Polubarinov spinor forms (up to a sign) part of a geometric object \emph{with (the conformal part of) the metric} overcomes Cartan's  ``\emph{insurmountable}'' \cite{CartanSpinor} difficulties in covariantly differentiating of spinors using the coordinate techniques of classical differential geometry.


 \subsection{Differentiating OP Spinors as (Almost Part of) Nonlinear Geometric Objects}

One often reads that spinors do not admit Lie differentiation, unless the generating vector field is a conformal Killing vector  field \cite{BennTucker} \cite[p. 101]{PenroseRindler2}.  
 Unlike other spinors, OP spinors do in fact have a classical Lie derivative, as one would expect from their having a spinorial coordinate transformation law and no additional gauge group.  Because the spinor forms only part of a geometric object (or something close enough), with the remainder being the symmetric square root of the metric $r_{\mu\nu}$ or the like, one in fact must consider a pair  such as   $\langle r_{\mu\nu}, \psi  \rangle$ as a candidate for Lie differentiation. Thus   $\langle r_{\mu\nu}, \psi,  \mathcal{L}_{\xi} r_{\mu\nu}, \mathcal{L}_{\xi} \psi \rangle$ is a geometric object (apart from coordinate restrictions and spinorial double-valuedness). 
Thus one obtains the desirable property that 
the commutator of two Lie derivatives is the Lie derivative with respect to the commutator of the generating vector fields.

Here two simplifications are possible by  making use of the theory of nonlinear geometric objects, which Ogievetsky and Polubarinov do not.  First,  $r_{\mu\nu}$ and $g_{\mu\nu}$ are equivalent as geometric objects (at least for admissible coordinate systems for 
$r_{\mu\nu}$)---that is, each is an algebraic function of the other, and  that function is the same in all coordinate systems.  One can therefore use the pair   $\langle g_{\mu\nu}, \psi  \rangle$  instead of   $\langle r_{\mu\nu}, \psi  \rangle$; while the spinor transformation law still depends on the metric and in a nonlinear way, at least the nonlinearity of the transformation rule for $r_{\mu\nu}$ is averted.  The fact remains that   $\langle g_{\mu\nu}, \psi  \rangle$ has a nonlinear transformation law because the spinor $\psi$ has a nonlinear metric-dependent transformation law, given in detail (at least infinitesimally) by OP \cite{OPspinorReprint}.  Thus a geometric object that contains the Lie derivative of the spinor field is
  $\langle g_{\mu\nu}, \psi,  \mathcal{L}_{\xi} g_{\mu\nu}, \mathcal{L}_{\xi} \psi \rangle$.  In fact $\psi$ is  redundant due to the linearity of transformation rule in $\psi$, so the relevant geometric object is merely 
$\langle g_{\mu\nu},   \mathcal{L}_{\xi} g_{\mu\nu}, \mathcal{L}_{\xi} \psi \rangle$. 
But the transformation rule for $\mathcal{L}_{\xi} \psi $ depends on $g_{\mu\nu}$ without depending on $  \mathcal{L}_{\xi} g_{\mu\nu}, $ so one finally has the geometric object $\langle g_{\mu\nu},   \mathcal{L}_{\xi} \psi \rangle$ for the Lie derivative of the spinor.  The second simplification will appear below.

Note that just as the covariant derivative of the metric with respect to the metric-compatible connection vanishes, the symmetric square root of the metric inherits the same property:  $\nabla g_{\mu\nu}=0  \leftrightarrow \nabla r_{\mu\nu}=0.$ One can see that result  by taking $$\nabla_{\alpha} g_{\mu\nu} = \nabla_{\alpha} ( r_{\mu\rho} \eta^{\rho\sigma} r_{\sigma\nu}) = 
 (\nabla_{\alpha}  r_{\mu\rho}) \eta^{\rho\sigma} r_{\sigma\nu} +   r_{\mu\rho} ( \nabla_{\alpha} \eta^{\rho\sigma}) r_{\sigma\nu} +  r_{\mu\rho} \eta^{\rho\sigma} \nabla_{\alpha} r_{\sigma\nu}$$ by the Leibniz rule, recalling that $\nabla_{\alpha} \eta^{\rho\sigma}=0$ because $\eta^{\rho\sigma}$ is a constant matrix, and noticing that the remaining terms give 40 equations related to 40 unknowns by a nonsingular linear transformation.  A perhaps  even more straightforward derivation that $\nabla r_{\mu\nu}=0$ uses the series expansion in equation (\ref{Series}) for $r_{\mu\rho}$ in terms of $g_{\mu\nu}$ and $\eta^{\rho\sigma}$ and notes that every factor in every term on the right side has vanishing covariant derivative.  Such a derivation applies, at least, whenever the series expansion holds.  But for any manifold, there is around any point a neighborhood where the series expansion converges. Even for nonlinear geometric objects, the vanishing of a covariant derivative is an invariant notion \cite{SzybiakCovariant}.  So this derivation by the series expansion works in general.

\section{Conformal \emph{In}variance of Dirac Operator with Densities: No Volume Element if  Massless }

A second simplification is possible, one not suggested by OP originally (but see \cite{OgievetskyLNC,BorisovOgievetskii}), but suggested by the conformal covariance of the Dirac equation \cite{InfeldSchildDirac,HitchinSpinor,ChoquetDeWitt2,BourguignonConference,Branson}. 
It is not often noted (except in  \cite{SchoutenHaantjesConformal,HaantjesConformalSpinor,PeresPolynomial} in four dimensions), but is quite true, that suitable use of tensor densities and densitized spinors can make the Dirac equation (absent a mass term) and its Lagrangian density manifestly conformally \emph{invariant}.
(It is mysterious that the MathSciNet review of (\cite{HaantjesConformalSpinor}) fails to notice the novelty of Haantjes's excluding the volume element using weighted spinors.) 
  The right choice of variables 
makes conformal transformations, which alter the volume element, do nothing at all to the variables actually used; the volume element disappears.  
The OP coordinate transformation rule for spinors does not depend on the volume element $\sqrt{-g},$ but only on the conformal part of the metric, $\hat{g}_{\mu\nu},$ a tensor density.  One can split the metric into irreducible pieces, those being the conformal part and the volume element \cite{PeresPolynomial,Anderson,Katanaev}, using the relation 
$g_{\mu\nu} = \hat{g}_{\mu\nu} \sqrt{-g} \, ^{ \frac{1}{2} };$ one sees that  $|\hat{g}_{\mu\nu}|=-1.$    This decomposition recently proved crucial in finding a counterexample to the Anderson-Friedman absolute objects program, in that GR itself turns out to have an absolute object in the volume element \cite{FriedmanJones,GiuliniAbsolute}.  
One can make a similar decomposition of the symmetric square root of the metric, or of its inverse $r^{\mu\nu}.$  To avoid the proliferation of symbols, I will express the determinant of the square root of the metric in terms of $g$; for the conformal part with anti-unit determinant, I again award a hat on top.  A relevant result is $r^{\mu\nu} = \hat{r}^{\mu\nu} \sqrt{-g} \, ^{-\frac{1}{4}}.$

While neither the elimination of the tetrad (saving 6 components) \cite{OPspinorReprint} nor the elimination of the volume element (saving 1 component)  \cite{SchoutenHaantjesConformal,HaantjesConformalSpinor,PeresPolynomial} is very new,  it is surprising how little known either procedure is.  Still more rare, and probably novel, is the combination of the two. The closest work that comes to mind
\cite{HamamotoNonlinear,NeemanSijackiAnnals,IvanenkoSardanashvily,NeemanSijackiBreakdown,BoulangerKirsch}  exponentiates a symmetric traceless entity, which many of the authors call ``shear'' by obvious analogy to 3-dimensional continuum mechanics, and get something like the square root of the conformal part of the metric. But getting that entity leaves unresolved whether it can be used, without $\sqrt{-g}$ in the picture, to write the massless Dirac equation. 

One payoff of using only the symmetric square root of the conformal metric density, rather than a tetrad, is that it is clear \emph{by inspection} that the symmetry group in (conformally) flat space-time is the 15-parameter conformal group.  With the tetrad legs present and supposedly needed, one might be reluctant to trust the local Lorentz invariance to conclude rigorously that only the metric, not the tetrad, must be considered.  With the volume element present, even with a transformation rule under conformal rescalings, it again takes at least some thought to realize that the volume element does not affect the symmetry group.  But with the surplus structures eliminated in favor of $\hat{r}_{\mu\nu},$ one needs only to use the fact that $\mathcal{L}_{\xi} \hat{r}_{\mu\nu}=0$ is equivalent to $\mathcal{L}_{\xi} \hat{g}_{\mu\nu}=0$ and the fact that the latter is the conformal Killing vector equation (with the surplus volume element purged), to infer that the 15-parameter conformal group is the group of symmetries of the non-variational object in the Lagrangian density in the conformally flat case.

 To avoid surplus structure,  one wants spinors, the square root of the metric, and densities together. The  Dirac operator $\gamma^{A} e_{A}^{\mu} \nabla_{\mu}$ that acts on spinors  is conformally covariant \cite{ChoquetDeWitt2,Branson}. One can show using densities (including densitized spinors) that there is a conformally \emph{invariant} Dirac operator  
 lurking in the vicinity; the appropriate spinor turns out to have weight $\frac{3}{8}$ in four space-time dimensions or, more generally, $\frac{n-1}{2n}$ in $n$ space-time dimensions. That conformally invariant Dirac operator is  $\gamma_{\mu} \hat{r}^{\mu\nu} \nabla_{\nu} \psi_{w},$ where $\gamma_{\mu}$ denotes a set of numerical Dirac matrices, $\hat{r}^{\mu\nu}$ is the symmetric square root of the inverse conformal metric density $\hat{g}^{\mu\nu},$  $\nabla_{\nu}$ is the Ogievetsky-Polubarinov covariant derivatives for spinors \cite{OPspinorReprint} with the density weight term (with the weight  altered to match the usual western rather than Russian  conventions), and $\psi_{w}$ is a spinor with weight $w = \frac{n-1}{2n}.$  No use is made of any scalar density in defining this operator, so it is a concomitant of just the weighted spinor and the conformal metric density. 
It turns out that the Lagrangian density in this formalism with the densitized variables is also manifestly conformally invariant in any dimension, because 
\begin{equation} \mathcal{L} = \sqrt{-g} \bar{\psi} \gamma_{\nu} r^{\nu\mu} \nabla_{\mu} \psi    =   \bar{\psi}_{w} \gamma_{\nu} \hat{r}^{\nu\mu} \nabla_{\mu} \psi_{w}. 
\end{equation}  
 $ \sqrt{-g}$ does not appear even algebraically in the latter expression. (Analogous results hold for the conformally covariant coupling of scalar fields (\emph{c.f.} \cite{Wald,ChoquetDeWitt2}): using scalar (density) fields with  weight $\frac{n-2}{2n},$  $ \sqrt{-g}$ does not appear in the Lagrangian density.) 
It is noteworthy that for (co)vector fields, taking the electromagnetic vector potential to be a weight 0 covector also achieves conformal invariance, at least in four dimensions.  One could, if one wished, raise the index and/or densitize the vector potential and  
spoil that invariance if one wished, leaving mere conformal covariance; but no one does so because the most naive case is also the nicest.   
 Much of the apparatus of conformal rescaling of fields \cite{Wald,ChoquetDeWitt2} can be avoided by the use of suitably weighted densities, which automatically depend on the metric in the appropriate way (or so one assumes initially, before taking the densities as primitive) and do not change under conformal rescalings.

%
%


\section{Differentiating OP Spinors with Conformal Invariance}

Ogievetsky and Polubarinov show that the Lie derivative for a spinor under coordinate transformations contains a new metric-dependent term, linear in the gradient of the generating vector field and highly nonlinear in $r^{\mu\nu}.$  If one introduces the decomposition $r^{\mu\nu} = \hat{r}^{\mu\nu} \sqrt{-g} \, ^{-\frac{1}{4}}$ into their equations 22 and 23 (of which the latter ought to have a factor of $\psi$ at the end), one can show that a change of integration variable from $\alpha$ to $ \alpha \sqrt{-g} \, ^{-\frac{1}{4}} $ makes $\sqrt{-g}$ disappear from their expression $\Delta$ and hence from the infinitesimal coordinate transformation formula and Lie derivative altogether.  This is not surprising: besides the conformal covariance, which we saw to admit reformulation to achieve conformal invariance,  
one also notices that the introduction of a local $O(1,3)$ group is served as effectively with a densitized tetrad $\hat{e}^{\mu}_A$ of weight $-\frac{1}{4}$; as a matrix, this tetrad has 15 independent components, the determinant being $1$ and the volume element having cancelled out. The local $O(1,3)$ (local Lorentz) index $A,$ not the coordinate transformation properties, is doing the work here, so altering the coordinate transformation properties by a choice of tensor density weight makes no difference.   Spinors, at least in the tetrad and OP formalisms, are deeply tied to the conformal part of the metric, but not  closely tied to the volume element.  Thus instead of an arbitrary nonsingular asymmetric 16-component $4 \times 4$ matrix of tetrad components, only a symmetric 9-component (anti-)unimodular matrix of components of the square root of the conformal metric density is required for the Dirac equation (apart from a mass term, which is trivial in its spinorial transformation properties).

Thus to form a geometric object (up to a sign and with time listed as the first coordinate, as usual) involving an OP spinor $\psi,$ one needs not the full metric tensor $g_{\mu\nu},$ but only the conformal metric density $\hat{g}_{\mu\nu}$, which of course is equivalent to $\hat{r}^{\mu\nu}.$  (One could view these as alternate coordinatizations of the same fiber.) 
Lack of  conformal \emph{in}variance  has led several authors to the mistaken conclusion that merely \emph{conformal} Killing vectors  are less special than are Killing vectors, as far as the massless Dirac equation is concerned   \cite{KosmannLie,KosmannLieApplied,FatibeneFermion,Cotaescu,FatibeneFrancaviglia}.  But that cannot be the case, because one cannot need the cooperation of an absent volume element any more than one can need the cooperation of the aether in Special Relativity.

 The entity  $\langle \hat{g}_{\mu\nu}, \psi \rangle$ still has a nonlinear transformation rule for $\psi,$ but gratuitous nonlinearity and the irrelevant volume element have been pared away.
Using the transformation rules for Lie and covariant derivatives for nonlinear geometric objects $\mathcal{L}_{\xi} \chi,$ $\nabla \chi,$ for $\chi$ \cite{SzybiakCovariant,SzybiakLie}, 
one has, schematically, for the components in the primed coordinate system in terms of those in the unprimed system,
 $(\mathcal{L}_{\xi} \psi) ^\prime \sim   \left( \frac{\partial \psi^\prime }{\partial \psi} \mathcal{L}_{\xi} \psi  + \frac{ \partial \psi^\prime }{\partial \hat{g}_{\mu\nu}} \mathcal{L}_{\xi} \hat{g}_{\mu\nu} \right)$.
One notices that $\mathcal{L}_{\xi} \hat{g}_{\mu\nu}=0$ is the equation for $\xi^{\mu}$ to be a conformal Killing vector field.  (Nowadays one usually  sees the conformal Killing equation written as $\mathcal{L}_{\xi} {g}_{\mu\nu} \sim {g}_{\mu\nu},$ thus introducing gratuitous reference to a volume element; the shoe of surplus structure again is on the other foot, that is, the modern foot, not the classical  foot.\footnote{What is the Lie derivative of an equivalence class? This formula gives an answer in a specific case.  Siwek gives a general answer to this question \cite{Siwek}, albeit one that he misapplies in the case of the object of Pensov \cite{SiwekCovariant} by neglecting the coordinate dependence of a certain scale in defining an equivalence relation. But why take the Lie derivative of an equivalence class when the geometric object $\hat{g}_{\mu\nu}$ is available?})  One sees that the Lie derivative with respect to conformal Killing vector is \emph{nicer}, because the second term 
$\frac{ \partial \psi^\prime }{\partial \hat{g}_{\mu\nu}} \mathcal{L}_{\xi} \hat{g}_{\mu\nu} $ then disappears.
But the Lie derivative exists for any vector field, not just conformal Killing vectors relative to the metric. It is evident that failure to include the (conformal part of the) metric with the spinor will lead to problems suggesting that spinors have Lie derivatives only for conformal Killing vector fields, just as one often hears \cite{PenroseRindler2,BennTucker}.  
 One thus explains a standard difficulty in spinor formalisms and overcomes it using the OP formalism.

For covariant derivatives, the situation is similar, but not identical:  
 $$(\nabla \psi) ^\prime \sim   \left( \frac{\partial \psi^\prime }{\partial \psi} \nabla \psi  + \frac{ \partial \psi^\prime }{\partial \hat{g}_{\mu\nu}} \nabla \hat{g}_{\mu\nu} \right).$$ 
Using the metric compatibility of the covariant derivative  (with the appropriate extra term for a density of weight $-\frac{1}{2}$ \cite{Schouten,Anderson,Israel})
 $$\nabla_{\alpha} \hat{g}_{\mu\nu}=         0,$$  
the unwanted new metric term disappears.  It therefore follows that 
 $\nabla \psi$ can be a  spinor-covector, as OP announce that it will be.   
 $\langle \hat{g}_{\mu\nu}, \nabla \psi \rangle$ is a geometric object.  $\psi$ itself does not appear due to the linearity \emph{in} $\psi$ of the transformation rule for $\psi$, while $\nabla \hat{g}_{\mu\nu}$ doesn't appear because it vanishes.    One could not impose that $\nabla \psi$  be a  spinor-covector without disposing of the new metric term---which would not be possible if an arbitrary connection were employed to take a covariant derivative. 


\section{Spinors, General Covariance, and Anderson-Friedman Absolute Objects}

An adequate account of general covariance must take spinor fields into account, because clearly there are electrons (or at any rate lumps of excitation of the electron field), and clearly in some (perhaps less fundamental) sense there are protons and neutrons as well, to say nothing of the less famous cousins posited by current physical theories and in many cases discovered in particle accelerators and the like.  Whereas Friedman's account of general covariance and absolute objects does not include spinors  \cite{FriedmanFoundations}, Anderson's book does include spinors.  Just how his treatment of spinors relates to his treatment of absolute objects is not terribly clear, however.  Although Anderson expected a correlation between whether a field counts as absolute or not (in terms of his criterion of gauge equivalence in all models) and whether it is varied in the least action principle to get Euler-Lagrange field equations, it is not clear what variational principle would apply.  Perhaps Heller and Bergmann's treatment \cite{BergmannHellerSpinor}, which Anderson did not mention but probably knew, would work, though its extra Lagrange multipliers are unappealing and savor of the irrelevant fields that Anderson's formalism bans  \cite{Anderson,TLL,LLN}.  More promising is the rather standard orthonormal tetrad formalism.

One major problem with the orthonormal tetrad formalism in the context of Anderson's absolute objects analysis of general covariance is that the tetrad includes a geometric object that is the same (up to gauge equivalence) in all models, an absolute object \cite{FriedmanJones}, which is just the sort of thing that substantively generally covariant theories are not supposed to have, according to Anderson.  Some time ago the Jones-Geroch dust counterexample was proposed as a counterexample to Anderson's analysis \cite{Jones,JonesGR,FriedmanFoundations}.
The 4-velocity of dust is a tangent vector that is nonzero wherever there is dust.  
 This proposed counterexample is incorrect 
\cite{FriedmanJones}, because an absolute object must be the same in all models (diffeomorphically equivalent between arbitrary neighborhoods), and dust has models with holes, yielding neighborhoods full of dust and neighborhoods that are empty.  But the mathematical fact that nonvanishing tangent vector fields are all alike up to diffeomorphisms remains.  (In fact there is a broader class of such objects, including tangent vector densities of any weight except $1$, first analyzed by Andrzej Zajtz \cite{ZajtzGerm} and later independently rediscovered by Robert Geroch in discussion with the author \cite{PittsPhilDiss}.) 
Whereas there possibly being holes in the dust makes the the Jones-Geroch dust example go away, the time-like leg of an orthonormal tetrad cannot ever vanish anywhere in any model.  It therefore is just the sort of counterexample that the Jones-Geroch dust 4-velocity field was believed to be.  If a tetrad is necessary for coupling spinor fields to a curved metric, then the existence of electrons generates a counterexample to Anderson's analysis of general covariance:  GR + electrons has an absolute object and hence violates Anderson's conception of substantive general covariance. A  tetrad leg, given coordinate and gauge freedom, can have components  $(1,0,0,0)$ in some neighborhood about any point.

The solution to this problem, as observed previously  \cite{FriedmanJones}, is to note that an orthonormal tetrad is not necessary for coupling spinor fields to a curved metric, because the OP spinor formalism lets spinors see the (square root of the) metric directly, without any orthonormal tetrad in the theory.  Six components of the 16 in the tetrad are irrelevant (leaving aside the volume element).  In the OP spinor formalism, these irrelevant bits are removed. Thus there is no tangent vector anywhere in the theory, once the surplus structure is removed.

As it happens, Geroch and independently Giulini have observed that the volume element $\sqrt{-g}$ in GR serves as a counterexample to Anderson's analysis of general covariance \cite{FriedmanJones,GiuliniAbsolute}.  Thus the viability of Anderson's analysis seems not to turn ultimately on a correct treatment of spinor fields. 
Evidently a correct analysis of substantive general covariance has not been found, or at any rate Anderson's rather attractive and widely known analysis is not it.  

But in fact the mild coordinate restrictions required by OP spinors implies that even the formal or trivial sort of general covariance, supposedly involving the admissibility of arbitrary coordinates, is problematic, contrary to widespread belief.  If my expectation that no one would bother to introduce a metric tensor in SR in order to admit swapping the coordinates $t$ and $x$ is accepted, then it is quite unclear why one should introduce a tetrad instead of just a metric---6 irrelevant field components---in order to remove the OP restriction that time be listed first.  


\section{OP Spinors, Conservation Laws, and Gravitational Energy Localization}

It has been generally thought necessary to treat spinorial energy-momentum in a very different fashion from the energy-momentum of fields that are tensors (or tensor densities, connections, or the like), on account of the absence of a Lie derivative for spinor fields for an arbitrary vector field \cite{MollerAnnals,FatibeneFermion,FatibeneFrancaviglia}. 
  But now that a Lie derivative for spinor fields is known, it follows that the bulk of the usual treatment of conserved quantities in terms of Lie differentiation also applies to spinor fields \cite{BilyalovConservation,EnergyGravity}.  There are a few technical complications arising from the nonlinearity of the (almost) geometric object $\langle g_{\mu\nu}, \psi \rangle,$
 but these are manageable and appropriate reflections of the character of spinors under coordinate transformations.  
It was shown recently that the usual objections to the localization of gravitational energy in terms of lack of a transformation law---\emph{i.e.}, failure to be a geometric object---do not withstand scrutiny \cite{EnergyGravity}.  The desire for such a transformation law can be justified only on the assumption that the conserved quantities  in different coordinate systems are equivalent, because that is just what transformation laws imply.  ``Physical meaning'' neither entails nor is entailed by a transformation law.  
But such equivalence is an arbitrary assumption.  There are infinitely inequivalent symmetries \cite{BergmannConservation}; why should one expect inequivalent symmetries to give equivalent conserved quantities? 
Why should one be bothered, for example, that the gravitational potential energy, involving the square of the coordinate gradient of the metric components, is highly nontrivial  for spherical coordinates in Minkowski space-time (Bauer's objection \cite{BauerEnergy,Pauli})?
Though `nothing happens' regarding the gravitational field, a great deal happens with the coordinates, making the relation between the two vary strongly with position; the Noether treatment entails that the pseudotensor involves a relation between the geometry and the coordinates. One can overcome the relation to a coordinate system by admitting all coordinate systems, obtaining a pseudotensor in all coordinate systems, an infinite-component entity \cite{EnergyGravity}. Such an entity is, in a technical sense of classical differential geometry, an ``object'' \cite[p. 28]{Nijenhuis} (in the sense of having a set of components at every point in every coordinate system about that point), though not a geometric object, due to its lack of a transformation law to make the components in different coordinate systems equivalent. 
 Christian M{\o}ller's fourth criterion, that the gravitational energy density be a scalar density for purely spatial coordinate transformations in order to make the energy in the laboratory room invariant \cite{MollerAnnals}, appears to be motivated only by wishful thinking or an analogy defeated by Noether's theorem.    Hence a treatment in terms of Noether's theorems and Lie differentiation faces no significant objections.  
With OP spinors admitting a Lie derivative, spinor fields can be included without conceptual difficulty.  Without OP spinors, one could not give such a unified treatment of energy-momentum localization and conservation in GR involving both bosonic and fermionic fields.


\section{Spinors and the Partial Conventionality of Simultaneity}

The conventionality of simultaneity has been a longstanding issue in the philosophy of physics \cite{ReichenbachSpace,GrunbaumSpace}.
The question has arisen whether spinor fields pose distinctive issues for the conventionality of simultaneity  \cite{ZangariSpinor,GunnSpinorSimultaneity,KarakostasSpinorSimultaneity,VetharianamTestSR,BainSpinorSimultaneity,JammerSimultaneity}. 
Zangari claimed that spinors require standard simultaneity and hence refute the thesis of the conventionality of simultaneity 
\cite{ZangariSpinor}.  (Here I do not have in mind questions of definability \cite{MalamentConvention}, but only the logically prior question of the range of coordinate freedom pertaining to simultaneity choices \cite{HavasSimultaneityConvention}.)

 Such a result would be a conceptual triumph if its technical foundation were correct, but   
Zangari's conclusions were refuted by Gunn and Vetharianam, who noted that Zangari had chosen an inappropriate representation of space-time coordinates and that an appropriate representation permits nonstandard synchrony conventions.  Gunn and Vetharianam employ a purpose-built generalization of the Dirac equation from the usual special relativistic form, without relating it to any standard formalism (though one can show that it has such relations).  Karakostas, while disagreeing on many technical points with Zangari and providing a more adequate technical basis, defends an anti-conventionalist thesis akin to Zangari's \cite{KarakostasSpinorSimultaneity}.  Part of Karakostas's work involves contact with the orthonormal tetrad formalism, according to which spinors transform as coordinate \emph{scalars}---that is, do not transform at all---and are spinorial with respect to an unrelated group that double-covers the local $O(1,3)$ group under which the local Lorentz indices (which I have generally written with capital Latin indices) transform.  One might expect the result of such a formalism, \emph{pace} Karakostas, to be that spinors \emph{as such} are irrelevant to the conventionality of simultaneity because spinors as such are coordinate scalars and hence do not even notice the choice of synchrony convention.

 Bain takes a nuanced position that leaves room for conventionalism  \cite{BainSpinorSimultaneity}. In particular, he refutes some of Karakostas's criticisms of Gunn and Vetharianam, but takes issue with their move that arguably trivializes the issue of spinors and conventionality by their breaking the link not simply between spinors and coordinates, but between spinors and space-time transformations of whatever sort.   From a mathematical standpoint, Gunn and Vetharianam's treating spinors as coordinate scalars---and presumably as spinors with respect to an unrelated $O(1,3)$ group---is unexceptionable.  (One readily sees that their formalism is a special case of the orthonormal tetrad formalism where one ignores the locality of the $O(1,3)$ group and refers only implicitly to a global $O(1,3)$ subgroup to identify $\psi$ as a spinor.  In standard coordinates their tetrad is simply the Kronecker $\delta,$ while their modified $\gamma$ matrices absorb the transformed tetrad.  Thus they pedagogically ignore all irrelevant complication to focus on the issue at hand.)   Bain's hesitation about their separating spinorhood from space-time stems from surprise at spin's being a key spacetime-related feature of elementary particles in Special Relativity and yet unrelated to spacetime (similar to the internal gauge symmetry of Yang-Mills fields) in General Relativity. Such a view of surprise was expressed long ago by Cartan himself:
\begin{quote}
Certain physicists regard spinors as entities which are, in a sense, unaffected by the rotations which classical geometric entities (vectors etc.) can undergo, and of which the components in a given reference system are susceptible of undergoing linear transformations which are in a sense autonomous.  See for example L. Infeld and B. L. van der Waerden \ldots.
It is clear that this impossibility [of having finite-component spinors with a covariant derivative in the classical Riemannian sense] provides an explanation of the point of view of L. Infeld and van der Waerden \ldots, which is however geometrically and even physically so startling. \cite[pp. 150, 151]{CartanSpinor}
\end{quote}
Detaching spinorhood from spatial coordinate rotations and the like was, to Cartan long ago as to Bain, quite a surprise--but an appropriate response to the impossibility of including spinors as such within the realm of coordinate transformations and covariant derivatives with Christoffel symbols, which Weyl and Cartan supposedly showed.  It is to Bain's credit that, at so late a date, he could still recover the surprise expressed by Cartan when the (supposed) impossibility was more novel.

 Due to the work of Ogievetsky and Polubarinov, that impossibility is overcome.  Cartan's and Bain's preference for unification, which they thought unsatisfiable,  is adequately realized using OP spinors, for which  the group that makes  $\psi$ a spinor is the (double cover of) a Lorentz group \emph{that is a subgroup of the space-time coordinate transformations}.  OP spinors, unlike the spinors of the tetrad formalism, are thus a strict generalization of the spinors in Minkowski space-time in Cartesian coordinates, such as one finds in standard particle physics books \cite{Kaku,PeskinSchroeder}.   While the answers regarding spinors and the conventionality of simultaneity have varied considerably, and all the previous answers have flaws, the question is a very welcome one in the sense that it provides a rare instance of consideration of the relevance of electrons, protons, and the like to philosophical questions about space-time theory.

OP spinors are  friendly to conventionalism about simultaneity, but not for reasons that have appeared previously. 
Anti-conventionalists Zangari and Karakostas have held that spinors \emph{as spinors} ought to be compatible with nonstandard simultaneity if a conventionalist thesis is to be upheld, and have denied that spinors as such are compatible with nonstandard simultaneity. Conventionalists Gunn and Vetharianam have thought it sufficient that spinors be present in a formalism admitting arbitrary coordinates, or at least coordinates arbitrary enough to admit nonstandard simultaneity conventions of the sort usually employed in this context.  That spinors \emph{as such} are coordinate scalars and hence wholly indifferent to simultaneity conventions is good enough for conventionalism, according to Gunn and Vetharianam. 
To Bain, such a move trivializes the anticonventionalist  point.  But in the OP formalism, which generalizes special relativistic theory as economically as possible (without the surplus structure of a tetrad and a new gauge group to deprive the extra fields of physical meaning), spinors do transform spinorially (not trivially) under coordinate transformations and do admit nonstandard simultaneity choices, as will appear below.  The anti-conventionalist's point is not trivialized, as Bain worried, but met head-on with  unfamiliar technical results.   A great variety of time coordinates is permitted, including all of them usually considered in the debate over the conventionality of simultaneity debate and some that are not.  That is not, however, because any coordinate whatsoever (nor even any `flat' one linearly related to standard coordinates in Minkowski space-time) is admissible as time; in fact some coordinates are inadmissible as time coordinates in the OP spinor formalism.  The conventional Kretschmannian wisdom is false (at least unless one inflates the formalism with 6 extra gauge compensation fields in the form of a tetrad).   The dividing line is apparently unprecedented in other formalisms, not corresponding, for example, to the Hilbert-M{\o}ller  proposal to make the time and space coordinates bear obvious temporal and spatial relationships to the null cone \cite{Hilbert2,Pauli,MollerBook}.  Instead of the nonlinear metric-independent transformation rule typical of OP spinors, for standard simultaneity a linear  metric-independent transformation rule arises, because transformations between standard coordinate systems lie within the 15-parameter conformal group for which linearity was proven above.

Let the transformation from standard to nonstandard simultaneity coordinates be given by $$x^{\mu^{\prime} } = (x^0 + (2\epsilon_1 -1)x^1 + (2\epsilon_2 -1) x^2 + (2\epsilon_3 -1)x^3, x^1, x^2, x^3).$$ One sees that standard simultaneity is the case $\epsilon_1=\epsilon_2=\epsilon_3=\frac{1}{2}.$  It is convenient to define $$\vec{n}=(2\epsilon_1 -1, 2\epsilon_2 -1, 2\epsilon_3 -1),$$ so then
$$x^{\mu^{\prime} } = x^{\mu} + \delta^{\mu}_0 n_i x^i.$$ Applying the tensor transformation rule to a rank 2 contravariant tensor, one finds the inverse metric for Minkowski space-time in with nonstandard simultaneity to be
\begin{eqnarray}
\left[
\begin{array}{cccc}
-1+\vec{n}^2  & n_1 & n_2 & n_3 \\ 
n_1  & 1 & 0 & 0 \\
n_2 & 0 & 1 & 0 \\
n_3 & 0 & 0 & 1 
\end{array}
\right]_.
\end{eqnarray}
These components have determinant $-1,$ so there is no difference between the metric and its conformal part in these coordinates. Let us calculate the symmetric square root of the metric $r^{\prime\mu\nu}$, which satisfies 
\begin{equation}
r^{\prime\mu\nu} \eta_{NA} r^{\prime\alpha\beta}=g^{\prime\mu\nu}.
\end{equation}
One way to do the calculation is to use the binomial series expansion \cite{OPspinorReprint}.  
Another way, which I will employ here, is to use a generalized eigenvalue formalism.  It is convenient to ignore two spatial dimensions and replace $\vec{n}$ with one number $n;$ there is no loss of generality, because one could rotate any $\vec{n}$ to be along the $x$-axis.  The generalized eigenvalue problem can be written as
\begin{eqnarray}
|g^{\prime\mu\nu} - \lambda \eta^{MN}| = \left| \left[
\begin{array}{cc}
-1+n^2 -\lambda  & n  \\ 
n  & 1 -\lambda  
\end{array}
\right] \right|  = -\lambda^2  + (2-n^2) \lambda -1=0.
\end{eqnarray}
One readily finds that 
\begin{eqnarray} 
\lambda = \frac{2 -n^2  \pm \sqrt{n^4 - 4 n^2}     }{2}.
\end{eqnarray}

Returning to the specific form of the eigenvalues, one sees that near the origin the eigenvalues tend to be complex, but farther from the origin ($|n| \geq 2$), the eigenvalues are real and \emph{negative}.  It follows that coordinates with $|n| \geq 2$ are inadmissible.  Coordinates with $|n| < 2$ are permitted.  
 The final result is
\begin{eqnarray}
r^{\prime\mu\nu}= 
\left[
\begin{array}{cc}
-\left(1-\frac{n^2}{2}\right) \left(1-\frac{n^2}{4}\right)^{-\frac{1}{2}}   & \frac{n}{2} \left(1-\frac{n^2}{4}\right)^{-\frac{1}{2}}  \\ 
  \frac{n}{2} \left(1-\frac{n^2}{4}\right)^{-\frac{1}{2}}&  \left(1-\frac{n^2}{4}\right)^{-\frac{1}{2}}
\end{array}
\right]  \nonumber \\%
 = %
\left[
\begin{array}{cc}
-1 + \frac{3n^2}{8}+\ldots    & \frac{n}{2} + \ldots  \\ 
  \frac{n}{2} + \ldots & 1+\frac{n^2}{8} + \ldots
\end{array}
\right]_.  
\end{eqnarray} 
Recalling that $n=2\epsilon -1$  and that the usual range of $\epsilon$ in discussions of the conventionality of simultaneity is between $0$ and $1$, it follows that the usual range of $n$ is between $-1$ and $1.$  Thus the Ogievetsky-Polubarinov formalism permits $-2 < n < 2,$ considerably larger than the typical $-1<n<1$ of the conventionality of simultaneity discussion, but less than the full  Kretschmannian arbitrariness expected based on experience with tensor calculus.  
The table compares the values discussed in the Reichenbachian conventionality of simultaneity literature, the values shown here to be admitted in the OP spinor formalism, and the values permitted by the Kretschmannian conventional wisdom that any coordinates are admissible.
\begin{center}
\begin{tabular}{| c | c | c |} \hline
 Reichenbachian   & Ogievetsky-Polubarinov & Kretschmannian  \\ 
  Simultaneity Conventionality & Spinor Formalism & Coordinate Arbitrariness  \\ \hline
$ 0 < \epsilon < 1$ & $-\frac{1}{2} < \epsilon < \frac{3}{2}$  &  any $\epsilon$         \\  \hline
$-1 < n < 1  $  &  $-2 <n<2$ & any $n$  \\ \hline
\end{tabular} 
\end{center}
Of course if one were prepared to interchange the order of the coordinates, one could admit any values for  $\epsilon$ and $n$ values in the OP formalism.  But such a move steps outside the traditional parameters of the conventionality of simultaneity discussion, which assumes that one already knows which coordinate is time.  

Recalling that the tetrad formalism involves making a coordinate transformation one the world (Greek) index and a  Lorentz transformation on the local (Latin) index, one can calculate the finite Lorentz transformation needed to symmetrize the tetrad.  The quantity 
\begin{eqnarray}
f_{\mu^{\prime} }^A r^{\prime\mu\nu} \eta_{NC}
\end{eqnarray}
is that Lorentz transformation.  For the Ogievetsky-Polubarinov formalism, one then applies that Lorentz transformation to the spinor field by multiplication by a suitably related matrix $S.$  After a few pages one finds that the spinor transformation $S$ is given by
\begin{eqnarray}
S= I \cosh \left[\frac{1}{2} \ln \sqrt{\frac{1-n/2}{1+n/2} }\right] + \gamma_0 \gamma^1 \sinh \left[\frac{1}{2} \ln \sqrt{\frac{1-n/2}{1+n/2} }\right]  \nonumber \\  %
= %
\frac{1}{2}I \left[   \sqrt[4]{\frac{1-n/2}{1+n/2} }  +    \sqrt[4]{\frac{1+n/2}{1-n/2} } \right]       + \frac{1}{2}\gamma_0 \gamma^1   \left[   \sqrt[4]{\frac{1-n/2}{1+n/2} }  -    \sqrt[4]{\frac{1+n/2}{1-n/2} } \right] \nonumber \\ %
= 
I - \frac{n}{4} \gamma_0 \gamma^1 + \ldots.    
\end{eqnarray}
(This exact expression confirms the result that $-2<n<2.$) In short, a coordinate transformation from standard simultaneity to nonstandard simultaneity induces an $n$-dependent  boost of the spinor.  The dependence on $n$ is inherited from the metric \emph{via} the symmetric square root of its conformal part.  The spinor transformation rule, one recalls, is linear in the spinor but nonlinear in the metric.  In this simple case the result is tractable for explicit calculations.


\section{The Schwarzschild Radius and  `Time' Coordinates}

Part of the lore of general relativists is the role of  ``Eddington-Finkelstein'' coordinates in the late 1950s in helping  to overcome `Schwarzschild singularity' at $r=2M;$ instead that radius came to be known as the ``horizon'' of a black hole, through which one might readily enough pass on the way to the  curvature singularity at $r=0$ \cite{FinkelsteinInfalling,MTW}. (Note that I am neither endorsing nor criticizing the details of this informal textbook history; the point at hand is philosophical.) Recalling such history, one might well conclude that holding too tightly to an association between a coordinate and some qualitative temporal or spatial character played a negative role in the context of discovery for the significance of $r=2M,$ a role only overcome with difficulty using infalling coordinates that made no such qualitative associations.\footnote{I thank Charles Misner for suggesting this sort of question.}    Does the  OP  formalism  un-learn lesson of Schwarzschild radius by re-regimenting coordinates as temporal and spatial? 

The infalling Eddington-Finkelstein coordinates are a radial coordinate $r$, a null coordinate $\tilde{V}$, and two angles $\theta,$ $\phi.$ The metric's line element is given by
$$ds^2 = -\left(1- \frac{2M}{r} \right) d\tilde{V}^2  + 2  d\tilde{V}dr + r^2(d\theta^2 + sin^2\theta d\phi^2)$$ \cite[p. 828]{MTW}. 
A null coordinate such as $\tilde{V}$ is, in some rough sense, half spatial and half temporal. $r$ seems quite unambiguously spatial.  One can use the generalized eigenvalue formalism to ascertain the coordinate ranges of admissibility for these coordinates.  The results are surprising in more than one respect.  One can show that the  OP admissibility range of $\langle \tilde{V}, r, \theta, \phi \rangle$ (noting the \emph{order}) is for $r> \frac{2M}{3}$. That seems plausible enough---something strange happens somewhere inside the Schwarzschild radius, but at least one can get inside it before having to switch coordinates.  That fact alone indicates that the OP `time' coordinate restrictions do not unlearn the lessons about coordinates from 1958 that contributed to the modern understanding of black holes.  At any rate one can get inside the horizon, and there is no reason to assume that  $r> \frac{2M}{3}$ is a real  barrier, so why not keep going with some other coordinate system?

One can also consider the  OP admissibility of $\langle r, \tilde{V}, \theta, \phi \rangle,$ with $r$ coming first. This is, intuitively, the `wrong' order, because a spatial coordinate is playing the role of OP `time,' while a null (half time, half space) coordinate is playing the role of space.  One might expect this coordinate system not to be admissible, or to be admissible only in some small exotic region, such as inside the horizon.  But on doing the calculation, one finds these coordinates to be admissible for  $r>0$!  The eigenvalues are complex.  As noted above, complex eigenvalues cause no trouble because they permit a real square root with eigenvalues in the right half of the complex plane.  Thus for  $r> \frac{2M}{3}$ there is more than one right order:  $\langle \tilde{V}, r, \theta, \phi \rangle$  and  $\langle r, \tilde{V},  \theta, \phi \rangle$ are both admissible.  The relation between the spinor components in the two systems is presumably quite nontrivial, in contrast to the relationship between the metric tensor components.  

In short, there are cases in which more than one right order exists, and cases where an intuitively wrong order is admissible and an intuitively more right order is inadmissible. What electrons know as `time' need not be unique or intuitive. Clearly OP spinors keep their own counsel in defining what counts as `time;' electrons don't so much know what a time coordinate is (conforming to some externally prescribed standard) as they \emph{stipulate} what a time coordinate is.  All of this follows from theorems in linear algebra about matrix square roots and negative eigenvalues, not from intuitions imported by hand.  The lesson of the  Schwarzschild radius is not unlearned by OP spinors.

\section{ Taking Stock}

No one, I think, will argue that special relativistic field theory for tensor fields is \emph{correct} when one permits arbitrary coordinates by using a metric tensor (under arbitrary transformations) and \emph{wrong} when one uses only Cartesian coordinates.  Neither will anyone argue that admitting arbitrary coordinates is generically more convenient, though it can be in specific applications.  Neither will anyone argue that admitting arbitrary coordinates is conceptually illuminating; certainly Peter Bergmann would not \cite{BergmannLectures}.  The same presumably holds for gauge symmetry in massive electromagnetism.  It isn't \emph{correct} to introduce Stueckelberg's gauge compensation field and incorrect to omit it and have no gauge freedom.  Neither is it generically more convenient to use Stueckelberg's trick, though in some applications it is.  Neither is it generically more illuminating to use it than to omit it.  If anything, the introduction of gauge compensation fields sheds conceptual darkness, not light, on theories, by tempting one to overlook the difference between merely formal gauge symmetries installed by a trick and substantive gauge symmetries characteristic of General Relativity,  Maxwell's (massless) electromagnetism, and (massless) Yang-Mills fields.  De-Ockhamization, such as setting `force' equal to the sum of `gorce' + `morce,' is, at least defeasibly,  not the path to conceptual enlightenment \cite{GlymourEpist,QuineEquivalent}.

Prior to the OP formalism, one did not know that the tetrad formalism was de-Ockhamized.  But now one has the Ockhamized formulation.  One thereby obtains greater unity between bosons and fermions within the theory of nonlinear geometric objects (up to a sign, with coordinate restrictions), a unified treatment of space-time symmetries in terms of Lie derivatives, and hence a unified understanding of conservation laws. Especially for conceptual questions, OP spinors are more illuminating than the tetrad spinor formalism.  The OP formalism  also merits a modern bundle treatment for the better treatment of global issues, matters not treated here.  Such treatment should  begin with work by Bilyalov.

Recalling that the Weyl-Cartan theorem about the need for an orthonormal tetrad depends heavily, for its significance at least, on failure to imagine the OP spinor formalism, one can imagine a counterfactual history, a rationally reconstructed one \cite{LakatosHistory}, one in which OP spinors are invented before Weyl and Cartan circulate their theorem.  It seems obvious that in such a world, the Weyl-Cartan spinor theorem, when invented, would be taken to imply the rather pedestrian conclusion that spinor fields require using nonlinear geometric objects (qualified by  coordinate restrictions and spinorial double-valuedness), not that tensor calculus (broadly construed) is inadequate or that an orthonormal basis is necessary.  While an orthonormal tetrad might be introduced at times for convenience, yielding a linear representation of a larger group rather than a nonlinear representation of a smaller group \cite{GatesGrisaruRocekSiegel}, no one would think the tetrad spinor formalism worthy of much philosophical contemplation.  De-Ockhamization can be convenient for technical purposes, but, at least without specific reason to think otherwise, it is not conceptually interesting.  The fact that introducing a tetrad would permit one the freedom to interchange the coordinates and thus obtain complete coordinate freedom would be thought far too trivial to justify the cost. It would be a way to fail to learn the lesson that the mathematics wants to teach, namely, that the signature makes a difference for spinors.  Plausibly, a rational process of updating beliefs in light of evidence should be independent of the order in which the pieces of evidence arise \cite{WagnerCommute,JeffreySubjective}, at least if some technical conditions hold.  The case at hand is nontrivial in that, like the problem of old evidence, it involves an expansion in logical-mathematical awareness, something that an ideal Bayesian agent could never experience.  Perhaps one can apply some strategies that have been used for the problem of old evidence \cite{GarberOldEvidence}. There might be an interesting methodological question here.  In any case, returning to the spinor example at hand, it seems evident that in a more rational world, OP spinors would be the default spinor formalism, especially for foundational purposes, while a tetrad might be introduced for convenience in some  practical calculations in cases where it exists.


\end{document}